\documentclass[review,12pt]{elsarticle}

\usepackage{amssymb}
\usepackage{amsmath}
\usepackage{physics}

\usepackage{booktabs}
\usepackage{subcaption}
\usepackage{multirow}

\usepackage{hyperref}

\usepackage[nameinlink]{cleveref}
\crefname{figure}{fig.}{Figs.}
\Crefname{figure}{Fig.}{Figs.}
\crefname{table}{table}{Tables}
\Crefname{table}{Table}{Tables}
\crefname{equation}{eq.}{Eqs.}
\Crefname{equation}{Eq.}{Eqs.}

\journal{Nucl. Instrum. Methods Phys. Res., Sect. A}

\begin{document}

\begin{frontmatter}



\title{ALPHANSO: Open-Source Modeling of ($\alpha$,n) Neutron Source Terms} 


\author[llnl,berkeley]{Divit Rawal\corref{cor1}}
\ead{divit.rawal@berkeley.edu}

\author[llnl]{Anthony J. Nelson\corref{cor1}}
\ead{nelson254@llnl.gov}
\author[llnl]{William Zywiec}
\author[llnl,berkeley]{Daniel Siefman}

\cortext[cor1]{Corresponding author}

\affiliation[llnl]{organization={Lawrence Livermore National Laboratory},
    city={Livermore},
    state={CA},
    postcode={94550},
    country={USA}}

\affiliation[berkeley]{organization={University of California, Berkeley},
    city={Berkeley},
    state={CA},
    postcode={94720},
    country={USA}}

\begin{abstract}
Applications ranging from nuclear safeguards to dark matter detection require accurate predictions of neutron yields and energy spectra produced by ($\alpha$,n) reactions. Legacy tools like SOURCES-4C remain widely used despite significant limitations, including outdated nuclear data, missing target nuclides, and restricted accessibility. Here, we present ALPHANSO, an open-source Python package for calculating ($\alpha$,n) neutron source terms. ALPHANSO incorporates modern nuclear data libraries and formats covering all naturally occurring target nuclides and provides a transparent, modular framework for updating or extending the data as new evaluations are released. Comparison with an updated version of SOURCES-4A, NeuCBOT, and experimental measurements across a range of elements and materials shows that ALPHANSO reproduces neutron yields and spectra in good agreement with experimental data and state-of-the-art ($\alpha$,n) calculations. These results demonstrate that ALPHANSO is a reliable, accessible, and modern alternative to legacy ($\alpha$,n) source term codes such as SOURCES-4C. Its open-source design and modular data handling make it readily extensible to future evaluated nuclear data and low-background applications.
\end{abstract}


\end{frontmatter}


\newcommand{\nuc}[2]{$^{#1}\mathrm{#2}$}


\section{Introduction}

Accurate modeling of neutron production from ($\alpha$,n) reactions is essential in a wide range of physics and engineering applications including nuclear safeguards \cite{SIMAKOV2017190}, waste management, criticality safety, astrophysics, reactor simulations, and the detection of rare events in the search for dark matter \cite{PhysRevD.98.102006}. Calculations of ($\alpha$,n) neutron source terms can be performed with explicit event-by-event Monte Carlo simulations or with deterministic methods \cite{GRIESHEIMER20171199, wrro238795}. Monte Carlo programs such as COG \cite{COG_manual_202590}, MCNP \cite{TechReport_2024_LANL_LA-UR-24-24602Rev.1_KuleszaAdamsEtAl}, or GEANT \cite{agostinelli_geant4simulation_2003} can simulate $\alpha$-particle transport and ($\alpha$,n) neutron production, but are computationally expensive. Deterministic codes such as SOURCES \cite{wilson_sources4c_2002}, in contrast, are capable of rapidly calculating source terms with low computational cost.

The most recent update to SOURCES is version 4C and it is one of the most widely used tools for calculating ($\alpha$,n) source terms. It is integrated into ORIGEN \cite{bell1973origen} and a version of it is shipped with MCNP \cite{bates2024mcnpsmesa}. SOURCES is a deterministic ($\alpha$,n) source term calculator originally developed at Los Alamos National Laboratory. It combines tabulated $\alpha$-decay data with ($\alpha$,n) cross sections and stopping power models to produce neutron spectra for user-defined materials and isotopic concentrations.

Despite its widespread use, SOURCES-4C suffers from significant limitations. It relies on obsolete nuclear data \cite{osti_1771892}. Its ($\alpha$,n) cross section libraries have not been updated since the mid-1980s. In addition, SOURCES data files do not conform to modern formats such as Generalized Nuclear Data Structure (GNDS) \cite{osti_23178677}, making updates difficult. As a result, improvements in evaluated nuclear data libraries cannot be easily incorporated. The code itself is also antiquated, being written in FORTRAN 77 in the early 1980s, and it has not been consistently maintained or validated since 2002.

SOURCES-4C is limited to $\alpha$-particle energies below 6.5 MeV. For the principal application of SOURCES-4C, the developers were primarily interested in ($\alpha$,n) reactions induced by actinide decay (mainly Am, Cm, Pu, U, Cf, and Bk), which have maximum $\alpha$-particle energies mostly below 6.1 MeV, so the imposed 6.5 MeV limit was reasonable for that scope. However, this excludes high-energy $\alpha$-particle emission energies from 21 nuclides in natural decay chains, which are critical for low-background studies and strongly impact dark matter experiments \cite{mei_evaluation_2009}. These shortcomings motivate the need for a modern ($\alpha$,n) source term calculator with updated data and flexible architecture.

Several previous efforts have attempted to overcome these limitations. NeuCBOT \cite{westerdale_radiogenic_2017} is a code written in Python in 2017. However, since it uses exclusively ($\alpha$,n) cross sections based on TALYS, it has lower accuracy than codes using evaluated libraries for some important target materials \cite{mendoza_neutron_2020}. We note that NeuCBOT is integrating JENDL support, but as this is not yet a released feature, we do not consider it here. Beyond the cross-section data, NeuCBOT has several additional limitations relative to ALPHANSO: it was originally written for Python~2.6 and is not designed to be imported as a Python library; it supports only the thick-target beam geometry (no interface or sandwich calculations); its decay-chain handling requires manually curated chain files rather than automatic computation of the $\alpha$ spectrum from an isotope inventory; and its cross-section data are stored in flat text files with no support for GNDS-formatted evaluated libraries. NEDIS \cite{vlaskin_neutron_2015} is a Russian code that supports $\alpha$-particles energies up to 9 MeV and has good accuracy. Unfortunately, this program is not publicly available outside of Russia. Other codes such as SaG4n \cite{mendoza_neutron_2020} use a biasing technique to increase the speed of Monte-Carlo methods (SaG4n in particular is based on Geant4), but are still significantly slower than ALPHANSO or other deterministic methods.\footnote{We find a $\sim$400$\times$ speedup between ALPHANSO and SaG4n.}

SOURCES-4A, an older version of SOURCES-4C, has been modified by researchers at the University of Sheffield to accept an improved nuclear data library, extending the energy range of incident $\alpha$-particles up to 10 MeV \cite{tomasello_calculation_2008}. They used the EMPIRE2.19 code  \cite{HERMAN20072655} to calculate the missing cross section data and transition probabilities, and the results are more accurate than those of SOURCES-4C. They have recently further improved the code through careful selection of default datasets \cite{Parvu2025Optimised}. Calculations made in this paper with the improved SOURCES-4A code use the most recent version, including the further refinements reported in 2025. We also note that the SOURCES codes require an additional reformatting step to incorporate new data, while ALPHANSO supports immediate drop in as new data is released.

To address the limitations of SOURCES-4C and related legacy codes, we present ALPHANSO (\underline{Alpha N}eutron \underline{So}urces)\footnote{ALPHANSO is publicly available on GitHub:~\href{https://github.com/alphanso-org/alphanso}{https://github.com/alphanso-org/alphanso}}, a modern, cross-platform Python package for calculating ($\alpha$,n) neutron source terms. ALPHANSO includes up-to-date evaluated nuclear data libraries (JENDL, TENDL, and ENDF) in the GNDS format~\cite{osti_23178677}. Users may also provide custom nuclear data, facilitating the integration of new evaluations. Its flexibility, reproducibility, and accessibility make ALPHANSO well-suited for modern nuclear data workflows and low-background experimental applications. ALPHANSO is distributed with a validation test suite that lets users directly evaluate the effect of updated or alternative nuclear data evaluations. 

The remainder of this paper is organized as follows: in \cref{sec:data} we give an overview of ALPHANSO's default data libraries and compare them to those of SOURCES. In \cref{sec:meth} we describe ALPHANSO’s calculation workflow and in \cref{sec:results} we compare its predictions against SOURCES, NeuCBOT and experimental data. Finally, in \cref{sec:conc} we summarize our findings and discuss ALPHANSO's role as a replacement for legacy ($\alpha$,n) source term codes.

\section{Data}\label{sec:data}

To calculate ($\alpha$,n) neutron source terms, ALPHANSO and SOURCES both require stopping power of $\alpha$-particles, level-branching fraction data, and ($\alpha$,n) cross sections.

\subsection{($\alpha$,n) Cross Sections}

For ($\alpha$,n) cross sections, ALPHANSO uses ENDF/B-VIII.1 \cite{nobre2025endfbviii1updatednuclearreaction}, JENDL-5 \cite{iwamoto_japanese_2023}, TENDL-2023 \cite{koning_tendl_2019} (generated with TALYS-1.96), and evaluated yields from Parvu et al. \cite{Parvu2025Optimised}. For each target nuclide, the data source was selected to maximize accuracy of neutron yield when compared to experimental data. For nuclides without reliable experimental neutron yield data, JENDL-5 was preferred where available and TENDL-2023 was used for nuclides for which no other data source exists.

ALPHANSO uses ENDF/B-VIII.1 for \nuc{6}{Li}, \nuc{9}{Be}, \nuc{17}{O}, and \nuc{18}{O}, for which it has comprehensive and recent evaluations. ENDF/B-VIII.1 specifically contains modifications of the JENDL-5 data for \nuc{9}{Be}, \nuc{16}{O}, and \nuc{17}{O} made by Naval Nuclear Laboratory \cite{osti_2571012}.

Parvu et al. provide updated, measurement-validated re-evaluations of light-element ($\alpha$,n) yields using a modified SOURCES-4A framework, correcting several known discrepancies in JENDL-5 for these nuclei, particularly in the $\alpha$-energy region relevant to U/Th decay chains. Their optimized cross sections reproduce thick-target neutron-yield benchmarks more accurately than JENDL-5 for C, Mg, Al, and Si, improving fidelity in homogeneous ($\alpha$,n) source calculations. ALPHANSO defaults to the Parvu et al. data for these target nuclides.

In comparison, SOURCES-4C uses experimental data collected at the time of release, and GNASH \cite{osti_6893289} calculations are used for isotopes without experimental data. Table~\ref{tab:data-sources} provides the data source for all target nuclides available in SOURCES-4A, as well as all targets for which JENDL data is available in ALPHANSO. In ALPHANSO, all stable nuclides not listed in this table use TENDL, generated with TALYS default parameters.

\begin{table}[h!]
  \centering
  \caption{Data sources for ($\alpha$,n) cross sections in ALPHANSO and SOURCES-4A~\cite{Parvu2025Optimised}.}
  \label{tab:data-sources}
  \begin{tabular}{lll}
  \hline
  \textbf{Nuclide} & \textbf{ALPHANSO} & \textbf{SOURCES-4A} \\
  \hline
  \nuc{6}{Li}   & ENDF/B-VIII.1 \cite{osti_2571012}         & JENDL-5 \cite{iwamoto_japanese_2023} \\
  \nuc{7}{Li}   & JENDL-5 \cite{iwamoto_japanese_2023}       & JENDL-5 \cite{iwamoto_japanese_2023} \\
  \nuc{9}{Be}   & ENDF/B-VIII.1 \cite{osti_2571012}         & Geiger \cite{Geiger1976} \\
  \nuc{10}{B}   & JENDL-5 \cite{iwamoto_japanese_2023}       & Prior + JENDL-5 \cite{iwamoto_japanese_2023} \\
  \nuc{11}{B}   & JENDL-5 \cite{iwamoto_japanese_2023}       & Prior + JENDL-5 \cite{iwamoto_japanese_2023} \\
  \nuc{13}{C}   & Parvu et al.\ \cite{Parvu2025Optimised}   & Parvu et al.\ \cite{Parvu2025Optimised} \\
  \nuc{14}{N}   & JENDL-5 \cite{iwamoto_japanese_2023}       & Gruhle (1972) + EMPIRE \cite{HERMAN20072655} \\
  \nuc{15}{N}   & JENDL-5 \cite{iwamoto_japanese_2023}       & EMPIRE 3.2.3 \cite{HERMAN20072655} \\
  \nuc{17}{O}   & ENDF/B-VIII.1 \cite{osti_2571012}         & Bair \cite{Bair1979} + JENDL-5 \cite{iwamoto_japanese_2023} \\
  \nuc{18}{O}   & ENDF/B-VIII.1 \cite{osti_2571012}         & JENDL-5 \cite{iwamoto_japanese_2023} \\
  \nuc{19}{F}   & JENDL-5 \cite{iwamoto_japanese_2023}       & Norman \cite{Norman1984} \\
  \nuc{21}{Ne}  & TENDL-2023 \cite{koning_tendl_2019}        & GNASH \cite{osti_6893289} \\
  \nuc{22}{Ne}  & TENDL-2023 \cite{koning_tendl_2019}        & GNASH \cite{osti_6893289} \\
  \nuc{23}{Na}  & JENDL-5 \cite{iwamoto_japanese_2023}       & JENDL-5 \cite{iwamoto_japanese_2023} \\
  \nuc{25}{Mg}  & Parvu et al.\ \cite{Parvu2025Optimised}   & Parvu et al.\ \cite{Parvu2025Optimised} \\
  \nuc{26}{Mg}  & Parvu et al.\ \cite{Parvu2025Optimised}   & Parvu et al.\ \cite{Parvu2025Optimised} \\
  \nuc{27}{Al}  & Parvu et al.\ \cite{Parvu2025Optimised}   & Parvu et al.\ \cite{Parvu2025Optimised} \\
  \nuc{29}{Si}  & Parvu et al.\ \cite{Parvu2025Optimised}   & Parvu et al.\ \cite{Parvu2025Optimised} \\
  \nuc{30}{Si}  & Parvu et al.\ \cite{Parvu2025Optimised}   & Parvu et al.\ \cite{Parvu2025Optimised} \\
  \nuc{31}{P}   & TENDL-2023 \cite{koning_tendl_2019}        & EMPIRE 3.2.3 \cite{HERMAN20072655} \\
  \nuc{37}{Cl}  & TENDL-2023 \cite{koning_tendl_2019}        & Woosley \cite{Woosley1975} \\
  \hline
  \end{tabular}
  \end{table}

\begin{figure}[htbp]
    \centering

    \begin{minipage}[t]{0.5\textwidth}
        \centering
        \includegraphics[width=\linewidth]{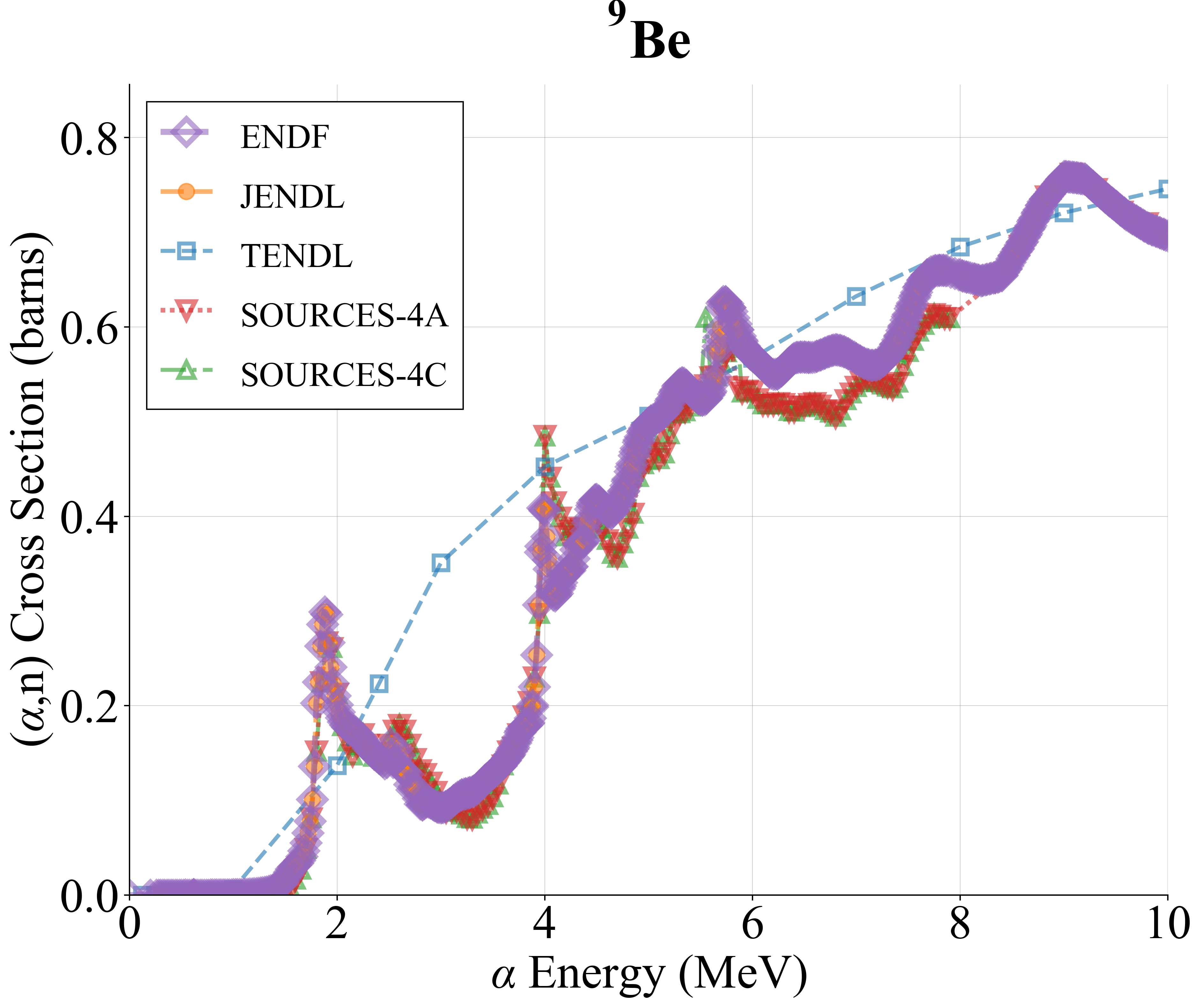}
    \end{minipage}\hfill
    \begin{minipage}[t]{0.5\textwidth}
        \centering
        \includegraphics[width=\linewidth]{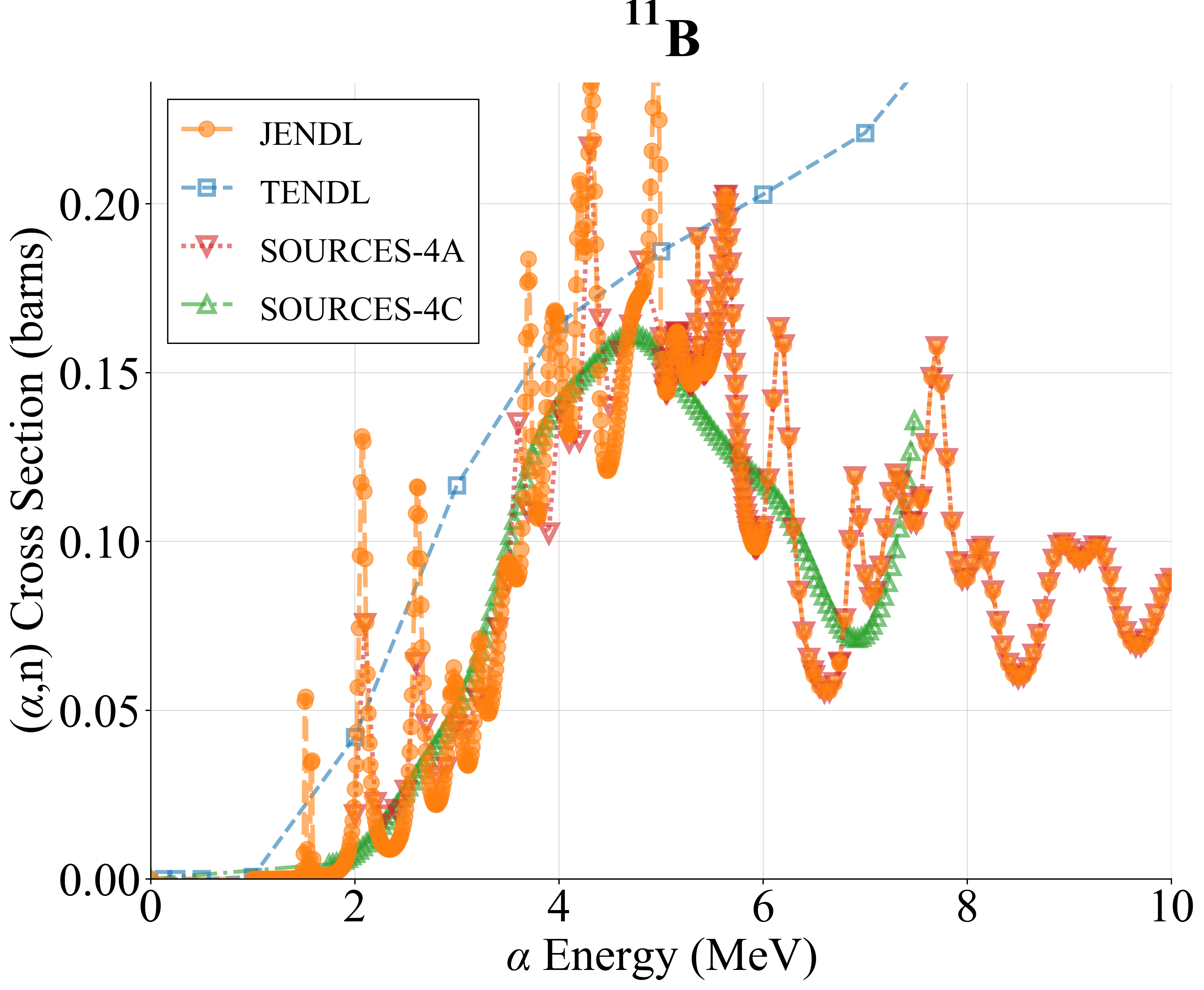}
    \end{minipage}

    \begin{minipage}[t]{0.5\textwidth}
        \centering
        \includegraphics[width=\linewidth]{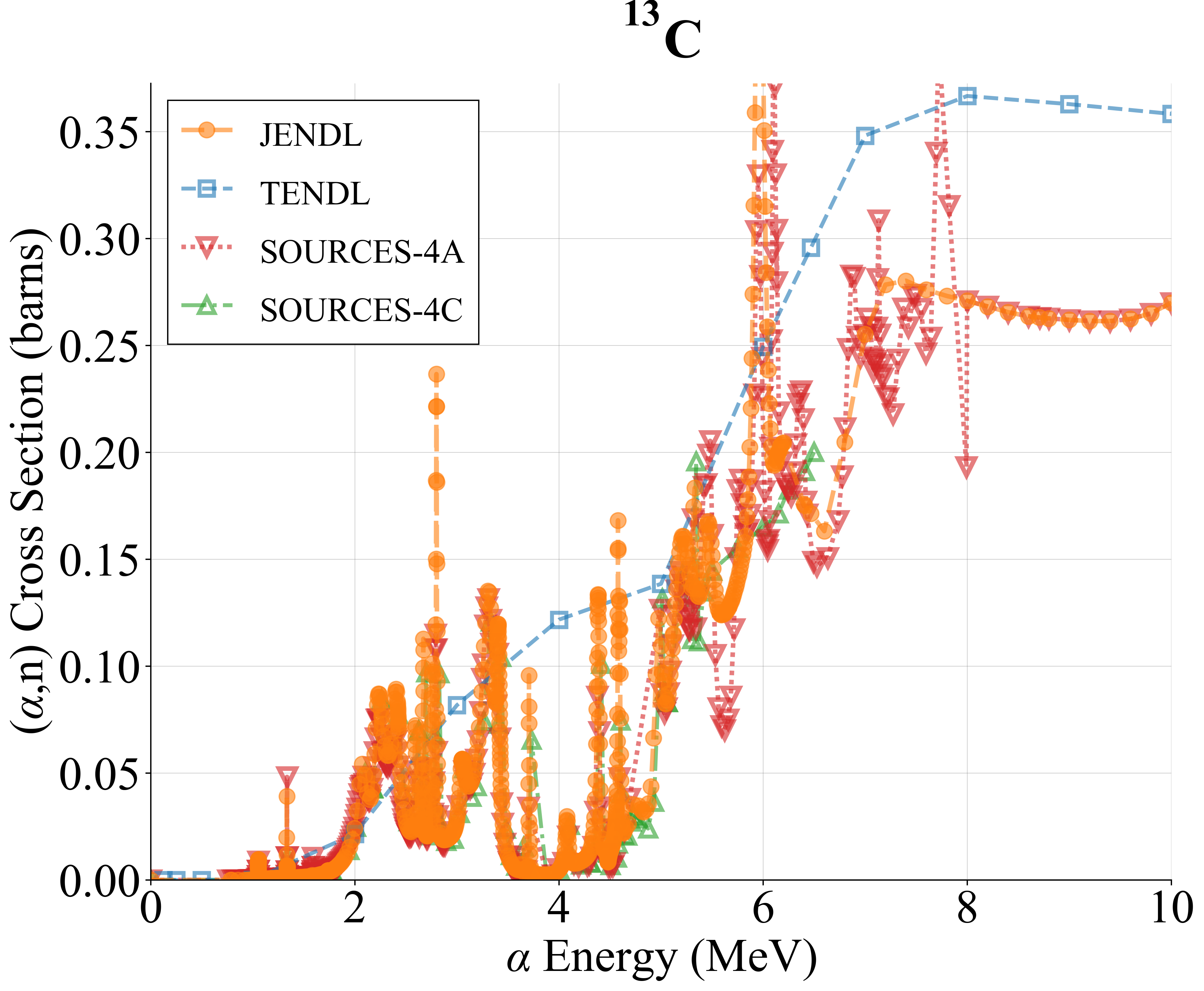}
    \end{minipage}\hfill
    \begin{minipage}[t]{0.5\textwidth}
        \centering
        \includegraphics[width=\linewidth]{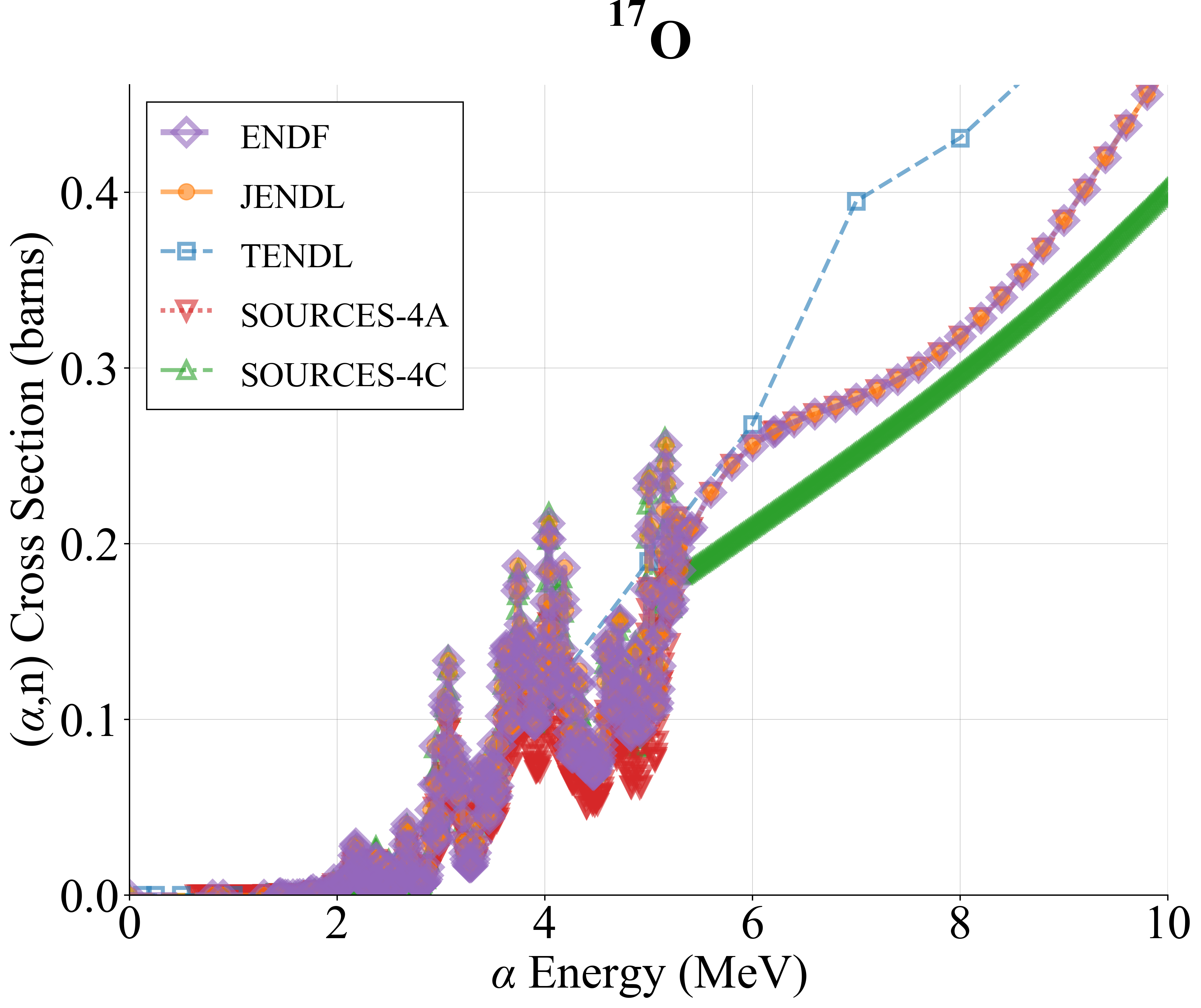}
    \end{minipage}

    \begin{minipage}[t]{0.5\textwidth}
        \centering
        \includegraphics[width=\linewidth]{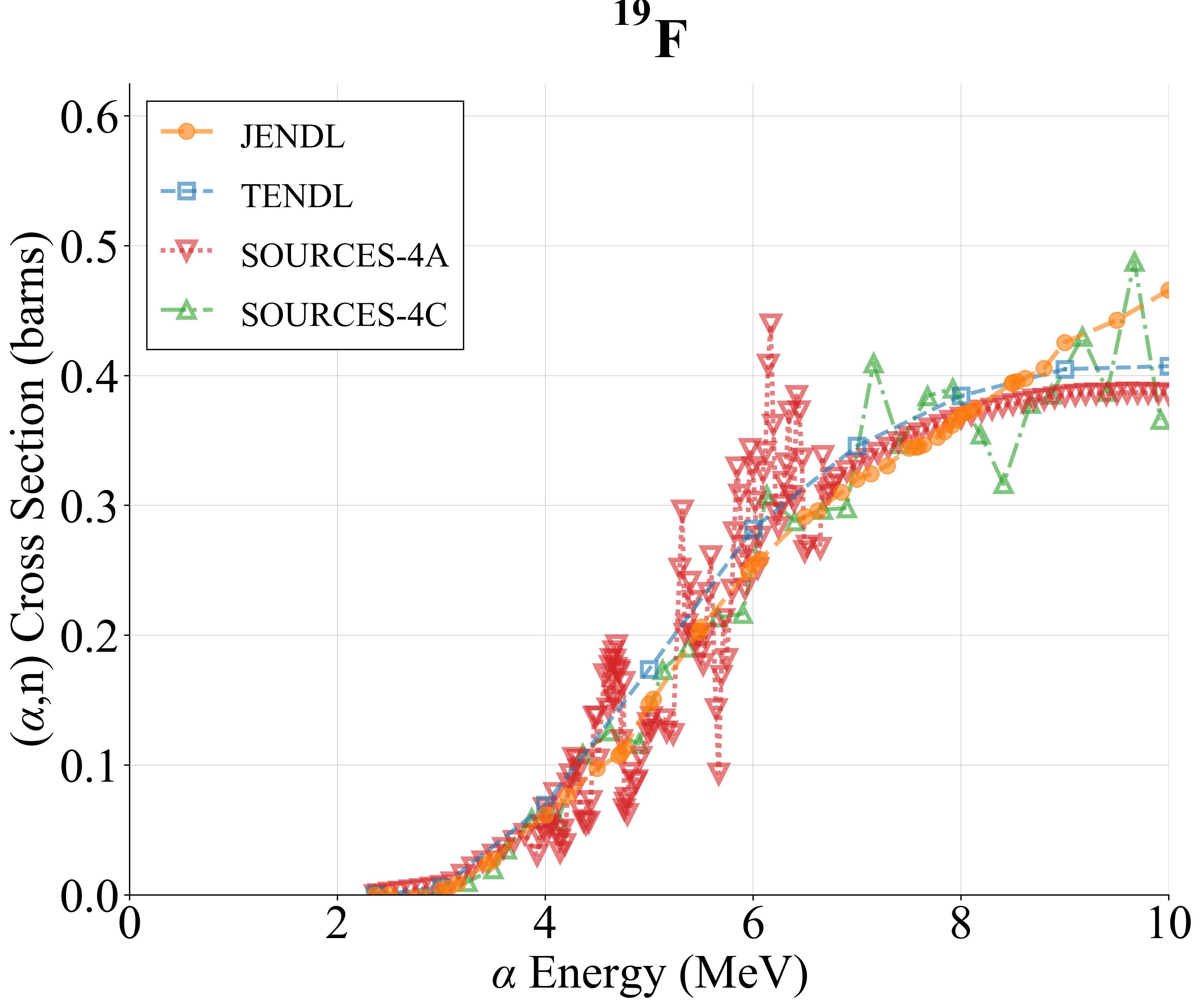}
    \end{minipage}\hfill
    \begin{minipage}[t]{0.5\textwidth}
        \centering
        \includegraphics[width=\linewidth]{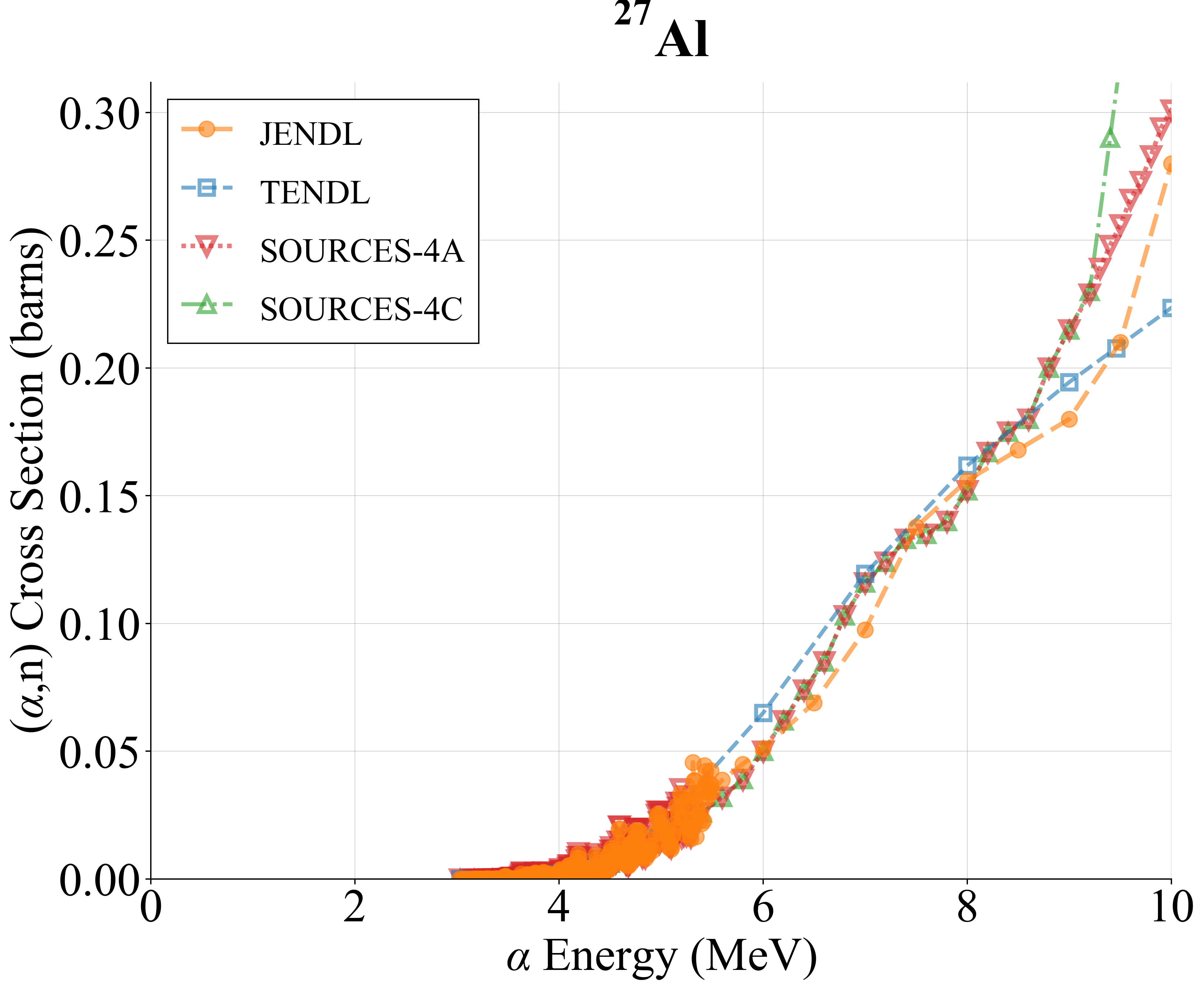}
    \end{minipage}

    \caption{($\alpha$,n) cross sections from ENDF/B-VIII.1, JENDL-5, TENDL-2023, SOURCES-4A, and SOURCES-4C for selected nuclides.}
    \label{fig:an_xs}
\end{figure}

A comparison between ENDF/B-VIII.1, JENDL-5, TENDL-2023, SOURCES-4A, and SOURCES-4C ($\alpha$,n) cross sections for several important target nuclides is provided in \cref{fig:an_xs}. As seen in the figure, TENDL-2023 data lack fine structure and are evaluated at fewer points than JENDL-5, ENDF/B-VIII.1, SOURCES-4A, or SOURCES-4C. This reflects a property of the TENDL-2023 distributed library itself: because TENDL provides pre-evaluated cross sections for thousands of nuclides, the ($\alpha$,n) data are tabulated on a coarse energy grid (typically $\sim$0.5 MeV steps) without the resonance structure that appears in more targeted evaluations such as JENDL-5.\footnote{ALPHANSO uses these library files directly via linear interpolation between tabulated points; we find that even using TALYS with a finer energy grid, the results are nearly identical.} Importantly, the discrepancy between TENDL-2023 and other evaluations is not only a resolution issue: even at energies where TENDL-2023 provides data points, the cross section magnitudes themselves differ significantly from JENDL-5, ENDF/B-VIII.1, and experimental benchmarks, reflecting the limitations of TALYS default nuclear model parameters for ($\alpha$,n) reactions at these energies. We note that these limitations may give inaccurate results, particularly for neutron emission spectra, and for this reason ALPHANSO defaults to JENDL-5 or ENDF/B-VIII.1 data where available, falling back to TENDL-2023 only for nuclides not covered by those libraries.

\subsection{Stopping Power}

\begin{figure}[htbp]
    \centering

    \begin{minipage}[t]{0.5\textwidth}
        \centering
        \includegraphics[width=\linewidth]{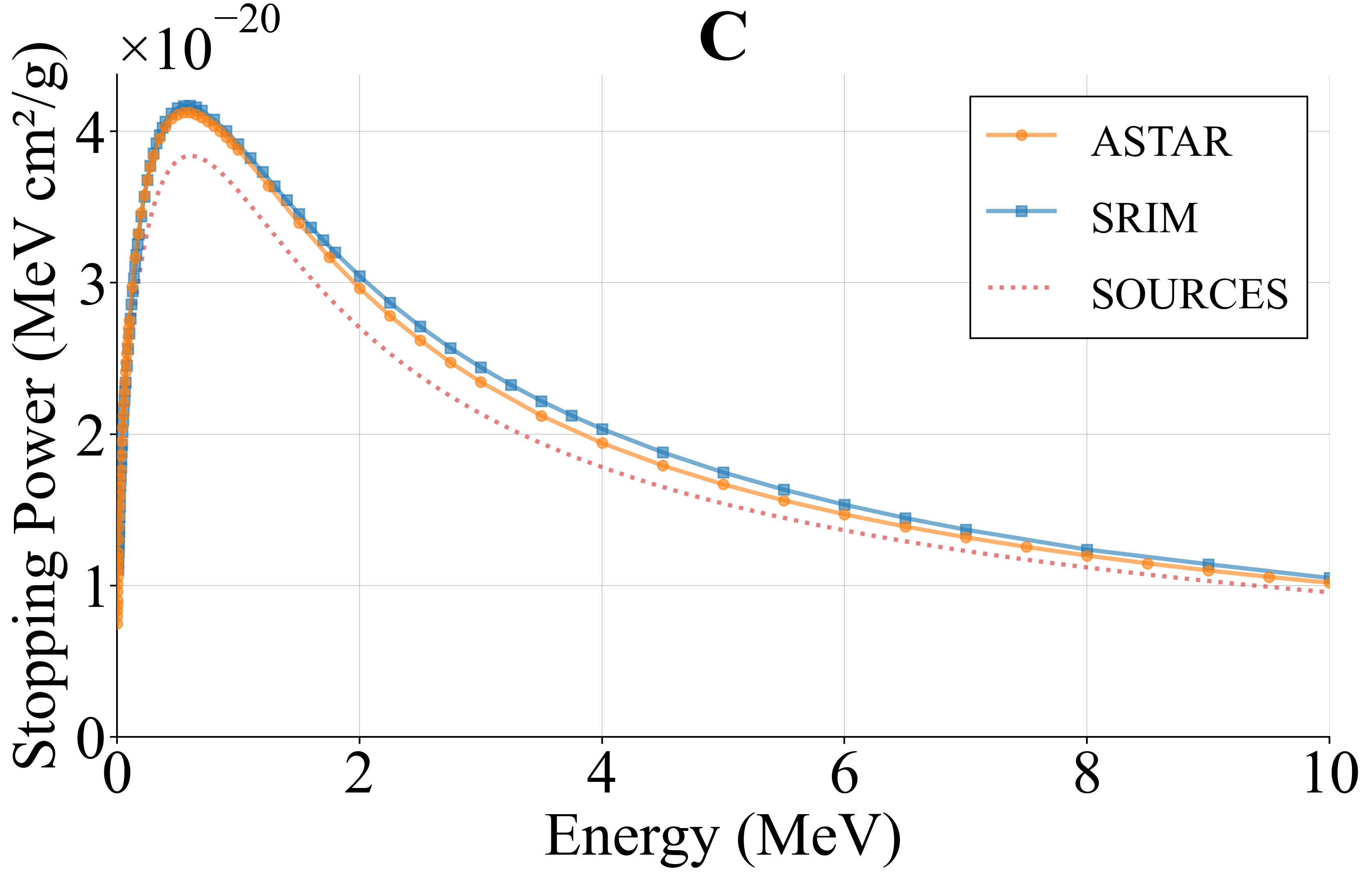}
    \end{minipage}\hfill
    \begin{minipage}[t]{0.5\textwidth}
        \centering
        \includegraphics[width=\linewidth]{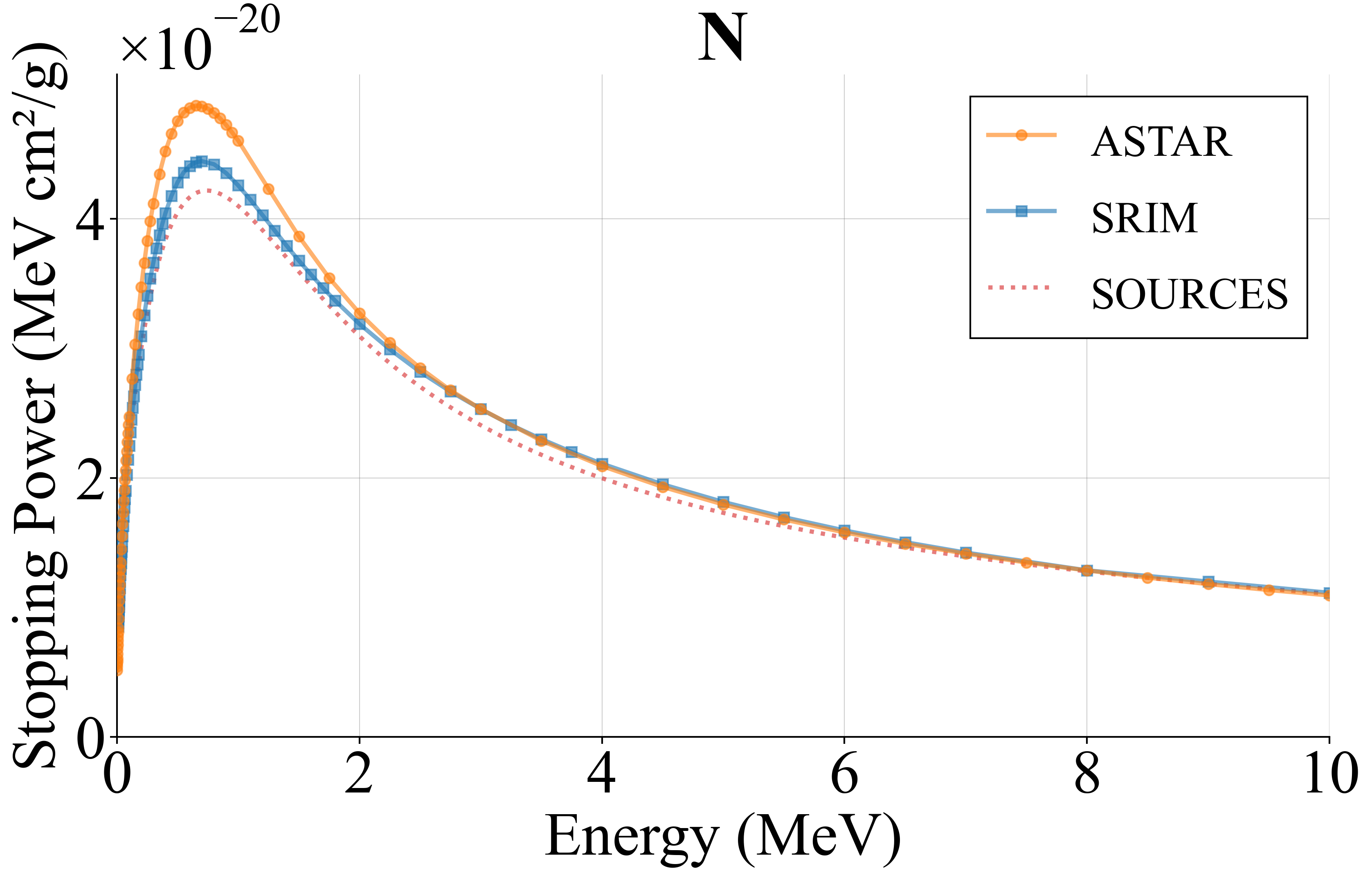}
    \end{minipage}

    \begin{minipage}[t]{0.5\textwidth}
        \centering
        \includegraphics[width=\linewidth]{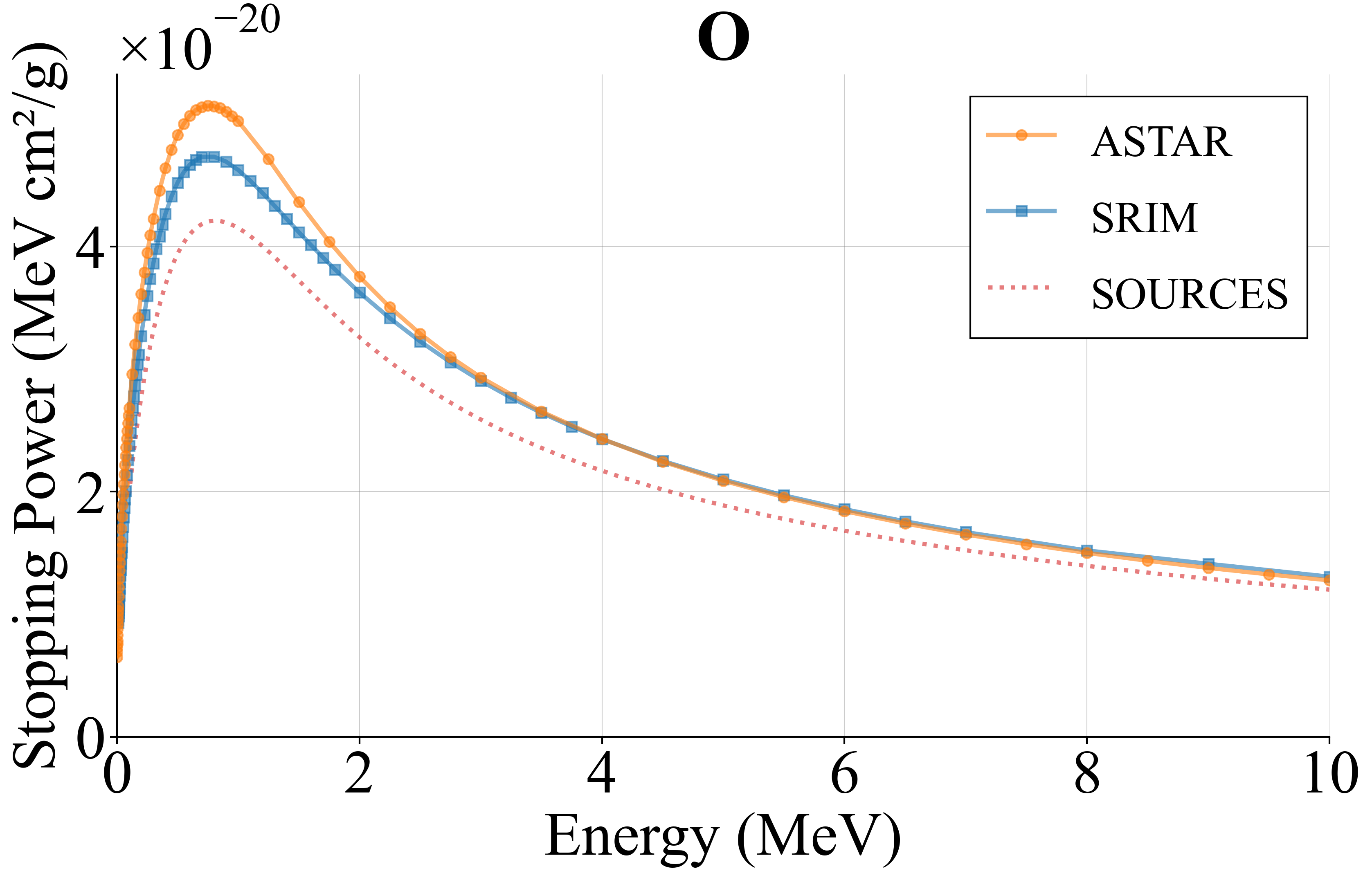}
    \end{minipage}\hfill
    \begin{minipage}[t]{0.5\textwidth}
        \centering
        \includegraphics[width=\linewidth]{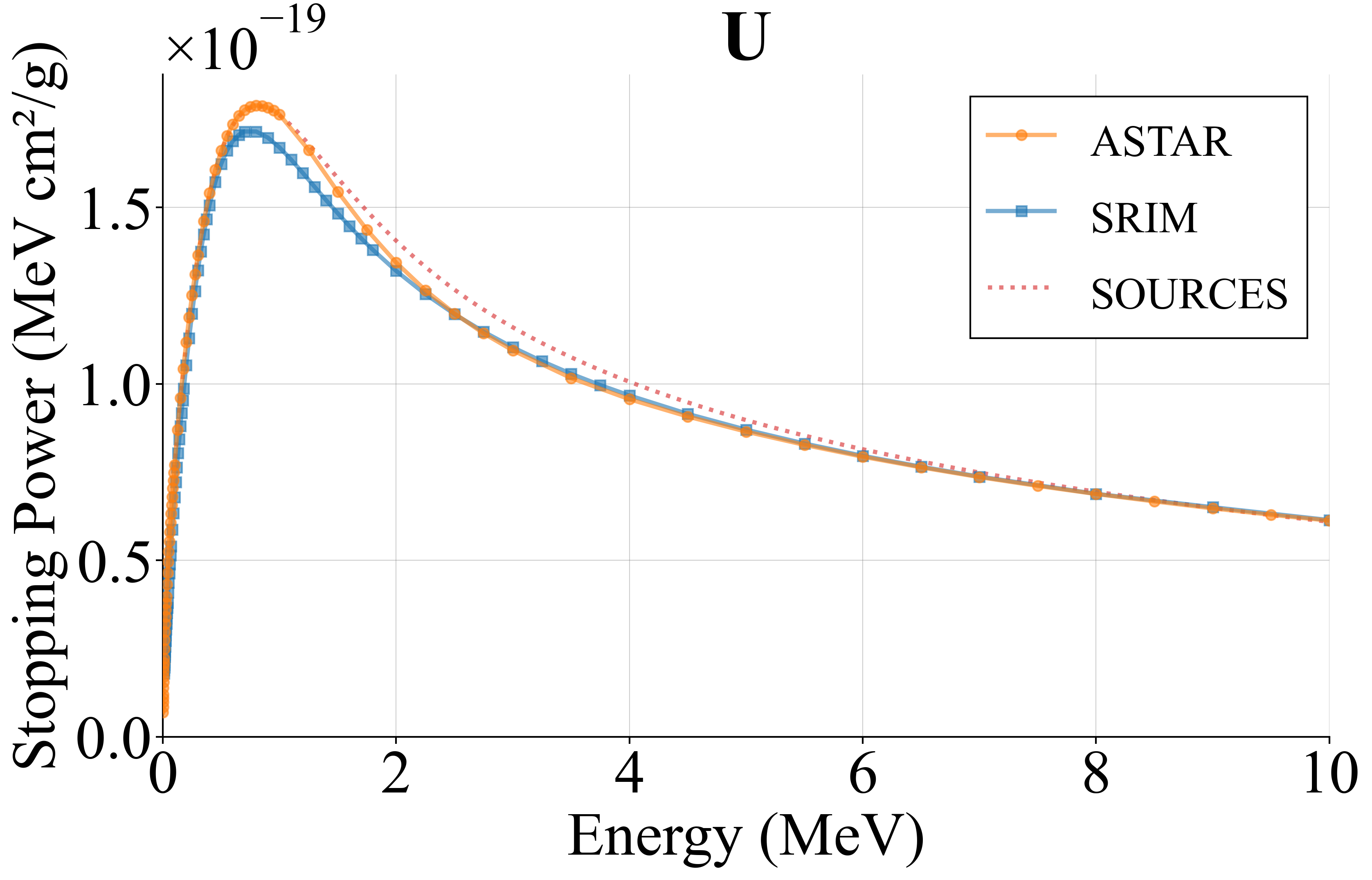}
    \end{minipage}

    \caption{Stopping power from ASTAR, SRIM, and SOURCES-4A/SOURCES-4C for selected elements. Note that SOURCES-4A and SOURCES-4C have identical stopping powers, plotted as ``SOURCES''.}
    \label{fig:stopping}
\end{figure}

ALPHANSO uses the ASTAR stopping power library \cite{berger_stopping-power_2009} where available, and otherwise uses data from SRIM \cite{ziegler_srim_2010} for elements with $Z \leq 92$. Elements available in ASTAR include H, He, Be, C, N, O, Ne, Al, Si, Ar, Ti, Fe, Cu, Ge, Kr, Mo, Ag, Sn, Xe, Gd, W, Pt, Au, Pb, and U. In SOURCES-4A and SOURCES-4C, stopping powers are taken from Ziegler et al.~\cite{ziegler_helium_1977} for elements with $Z \leq 92$. ALPHANSO and SOURCES both use stopping power coefficients from Perry and Wilson \cite{Perry1981} for $92 < Z \leq 105$. These coefficients are employed to reconstruct the Ziegler-Biersack-Littmark (ZBL) universal stopping power fit. Both SOURCES and ALPHANSO represent stopping power as a function of atomic number only. A comparison of stopping powers from SOURCES, ASTAR, and SRIM is shown in \cref{fig:stopping}. Generally, the data sources show largest disagreement below 1 MeV. Importantly, the neutron production cross section generally becomes very small in that energy range. That low cross section de-emphasizes the importance of the difference in stopping powers on integral quantities like total ($\alpha$,n) yield and neutron spectra.

\section{Methodology}\label{sec:meth}

ALPHANSO and SOURCES follow the same thick-target, continuous slowing down approximation physics model for calculating both the neutron yield and the emitted neutron energy spectrum from ($\alpha$,n) reactions. A fundamental assumption is that the targets are infinitely thick, i.e., all $\alpha$-particles stop in the target for the energy ranges under consideration. For each nuclide $i$ in a material, the neutron yield $Y_i$ from an $\alpha$-particle with initial energy $E_\alpha$ is given by \cref{eq:Yi}:

\begin{equation}
    Y_i(E_\alpha) = \frac{\eta_i}{\eta} \int_{0}^{E_\alpha} \frac{\sigma_i^{(\alpha, \text{x}n)}(E)}{\abs{\varepsilon(E)}}\ \mathrm{d}E,\label{eq:Yi}
\end{equation}

where $\eta_i$ is the number density of atoms of type $i$ in the material, $\eta$ is the total number density of all atoms in the material, $\sigma_i^{(\alpha, \text{x}n)}(E)$ is the neutron production cross section for nuclide $i$ (including all relevant channels), and $\varepsilon(E) \doteq -\frac{1}{\eta}\frac{\mathrm{d}E}{\mathrm{d}x}$ is the (per-atom) stopping power of the material (which may include many nuclides in different concentrations). In ALPHANSO, the integral \cref{eq:Yi} is computed via trapezoidal integration. For materials composed of $J$ elemental constituents, the stopping power $\varepsilon(E)$ of the material is computed via the Bragg-Kleeman \cite{bragg__1905} relationship (\cref{eq:bk})

\begin{equation}
    \varepsilon(E) = \frac{1}{\eta} \sum_{j=1}^J \eta_j \varepsilon_j(E),\label{eq:bk}
\end{equation}
where the total number density $\eta = \sum_{j=1}^J \eta_j$.

\Cref{eq:Yi} results from integrating the reaction probability along the slowing-down path and changing variables from path length to energy via $\mathrm{d}x = -\frac{\eta}{\varepsilon(E)}\,\mathrm{d}E$. \Cref{eq:YE} gives the total yield as a sum over all target nuclides present:

\begin{equation}
  Y(E_\alpha)=\sum_i Y_i(E_\alpha).\label{eq:YE}
\end{equation}

For sources that have multiple $\alpha$-particle emission energies $E_{\alpha,k}$ with intensities $I_k$ (normalized to $\sum_k I_k=1$), ALPHANSO and SOURCES compute $Y(E_{\alpha,k})$ for each line and report the intensity-weighted yield $\sum_k I_k\,Y(E_{\alpha,k})$.

The neutron energy spectrum is constructed by combining the above yield with two-body kinematics, under the assumption of isotropic emission of neutrons in the center-of-mass frame. For each reaction channel that produces a residual nucleus in excitation state $m$, \cref{eq:Q} gives the effective $Q$-value.

\begin{equation}
    Q_m = Q - E_\text{ex},\label{eq:Q}
\end{equation}

where $Q$ is the ground-state reaction $Q$-value and $E_\text{ex}$ is the excitation energy of the residual nucleus. \Cref{eq:mr} defines the dimensionless mass ratios:

\begin{equation}\label{eq:mr}
    a_1 \doteq \frac{m_n}{m_\alpha}, \qquad
    a_2 \doteq \frac{m_t}{m_\alpha}, \qquad
    a_3 \doteq \frac{m_n}{m_r},
\end{equation}

with $m_n$ the neutron mass, $m_\alpha$ the $\alpha$-particle mass, $m_t$ the target mass, and $m_r$ the residual nucleus mass. Using these definitions and conservation of momentum and energy, the kinematic bounds of the emitted neutron energy $E_n$ (where the subscript $n$ denotes the outgoing neutron) in the lab frame for target nuclide $i$ in excitation state $m$ are given in \cref{eq:Epm}:

\begin{equation}\label{eq:Epm}
    E^{\pm}_{n,i,m} = \left(\sqrt{\frac{E_\alpha a_1}{1+a_2}} \pm \sqrt{\frac{Q_{i,m}}{1+a_3} + \frac{E_\alpha a_2}{(1+a_2)(1+a_3)}} \right)^2.
\end{equation}

Each open reaction channel $(i,m)$ --- where $i$ again identifies the target nuclide and $m$ the residual nucleus excitation state, as above --- is further characterized by a branching ratio $b_{i,m}(E_\alpha)$, corresponding to the probability that an $\alpha$ of energy $E_\alpha$ incident on nuclide $i$ produces a residual nucleus in excitation state $m$. These branching ratios are derived from the individual level cross sections (MT=50--59) in the same evaluated nuclear data files used for the total ($\alpha$,n) cross sections (ENDF/B-VIII.1, JENDL-5, TENDL-2023, or SOURCES-4C, following the source hierarchy in Table~\ref{tab:data-sources}). When level cross-section data are absent, a branching ratio of unity to the ground state is assumed. These branching ratios provide the relative weighting among the $Q_m$ values. Therefore the channel-specific yield (\cref{eq:cy}) is

\begin{equation}\label{eq:cy}
    Y_{i,m}(E_\alpha) = b_{i,m}(E_\alpha)Y_i(E_\alpha).
\end{equation}

Assuming a uniform neutron energy distribution between $E_{n,m}^-$ and $E_{n,m}^+$, the differential yield spectrum is given by \cref{eq:dy}:

\begin{equation}\label{eq:dy}
    \frac{\mathrm{d}Y}{\mathrm{d}E_n}(E_\alpha) \approx
    \sum_{i,m}
    \frac{Y_{i,m}(E_\alpha)}{E^+_{n,i,m}-E^-_{n,i,m}}
    \mathbf{1}_{\{E^-_{n,i,m} \leq E_n \leq\,E^+_{n,i,m}\}},
\end{equation}

where $\mathbf{1}_{\{E^-_{n,i,m} \leq E_n \leq\,E^+_{n,i,m}\}}$ is the indicator function that equals 1 if $E_n$ lies within the kinematically allowed range for nuclide $i$ and excitation state $m$, and 0 otherwise.

Finally, for a source that has multiple $\alpha$ emission energies $E_{\alpha,k}$ with intensities $I_k$ (normalized such that $\sum_k I_k = 1$), ALPHANSO computes the intensity-weighted spectrum as in \cref{eq:iws}:

\begin{equation}\label{eq:iws}
    \frac{\mathrm{d}Y}{\mathrm{d}E_n} =
    \sum_{k} I_k \frac{\mathrm{d}Y}{\mathrm{d}E_n}(E_{\alpha,k}).
\end{equation}

\subsection{Software Interface and Supported Calculation Types}\label{subsec:interface}

ALPHANSO accepts inputs via a YAML configuration file or a Python dictionary, and can be invoked from the command line (\texttt{alphanso config.yaml}) or integrated directly into Python workflows via \texttt{Transport.calculate(config)}. The primary inputs are a material composition---specified as a dictionary of isotope or element identifiers with associated mass fractions---and the problem geometry type. Optional inputs include custom nuclear data directories, allowing users to substitute alternative cross-section or stopping-power datasets.

ALPHANSO supports four calculation geometries, each targeting a distinct physical scenario. A beam calculation computes the neutron yield and spectrum for a mono- or poly-energetic $\alpha$-particle beam incident on a thick homogeneous target; this is the geometry used for comparisons with accelerator experiments in \cref{sec:results}. A homogeneous calculation treats a uniform mixture of $\alpha$-emitting and target isotopes (e.g., a plutonium oxide pellet), computing the combined ($\alpha$,n) and spontaneous-fission neutron yield in units of n/s/g; spontaneous-fission contributions are modeled using the Watt fission spectrum, with parameters drawn from ENDF/B-VIII decay data, and the resulting SF and ($\alpha$,n) spectra are reported both separately and combined. An interface calculation handles a planar boundary between a pure $\alpha$-source region and a separate target region, returning the neutron yield per unit interface area (n/s/cm$^2$). A sandwich calculation extends the interface geometry to include one or more intermediate material layers between the source and target regions, resolving per-layer yield contributions.

For all geometry types, ALPHANSO outputs the integrated neutron yield and a binned neutron energy spectrum, both normalized and absolute. When an output directory is specified, results are written to YAML files for reproducibility. Results are reported at the material level; contributions from individual $\alpha$-emitting or target nuclides and individual excitation levels of the residual nucleus are summed into the aggregate spectrum, except that for homogeneous calculations the scalar spontaneous-fission yield and Watt parameters are reported separately per SF nuclide via the \texttt{sf\_contributors} field. A validation test suite is distributed with the package, enabling users to directly assess the impact of alternative nuclear data evaluations.


\section{Results}\label{sec:results}

To validate ALPHANSO, we compare its neutron yield and energy spectra against both experimental data \cite{west_measurements_1982,jacobs_energy_1983} and calculations from SOURCES-4C and SOURCES-4A. We also provide a brief analysis of the effect of data source choices in \cref{subsec:data_source_analysis}.

\subsection{Monoenergetic Beam Yield}

\begin{figure}[!h]
    \centering

    \begin{minipage}[t]{0.5\textwidth}
        \centering
        \includegraphics[width=\linewidth]{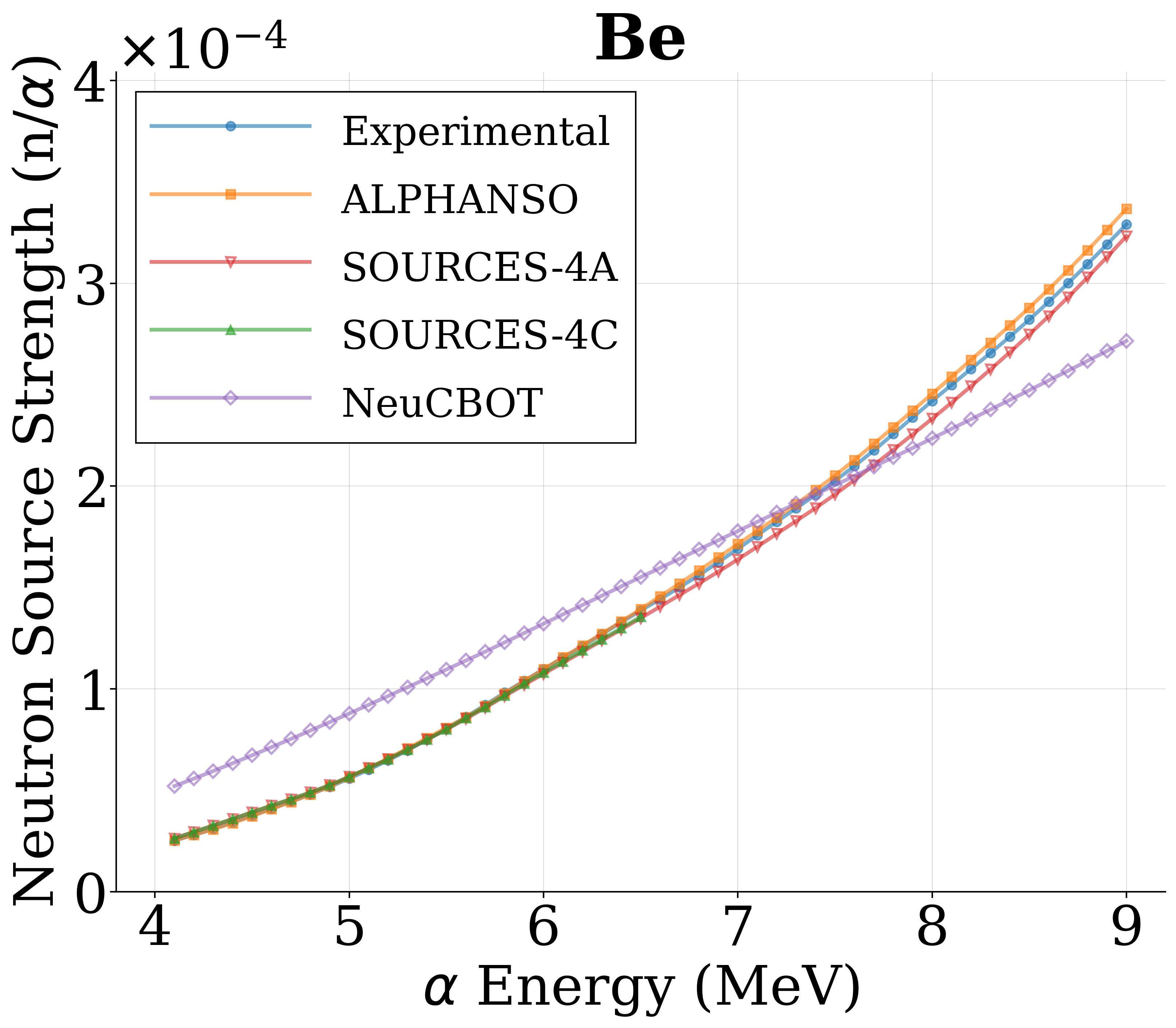}
    \end{minipage}\hfill
    \begin{minipage}[t]{0.5\textwidth}
        \centering
        \includegraphics[width=\linewidth]{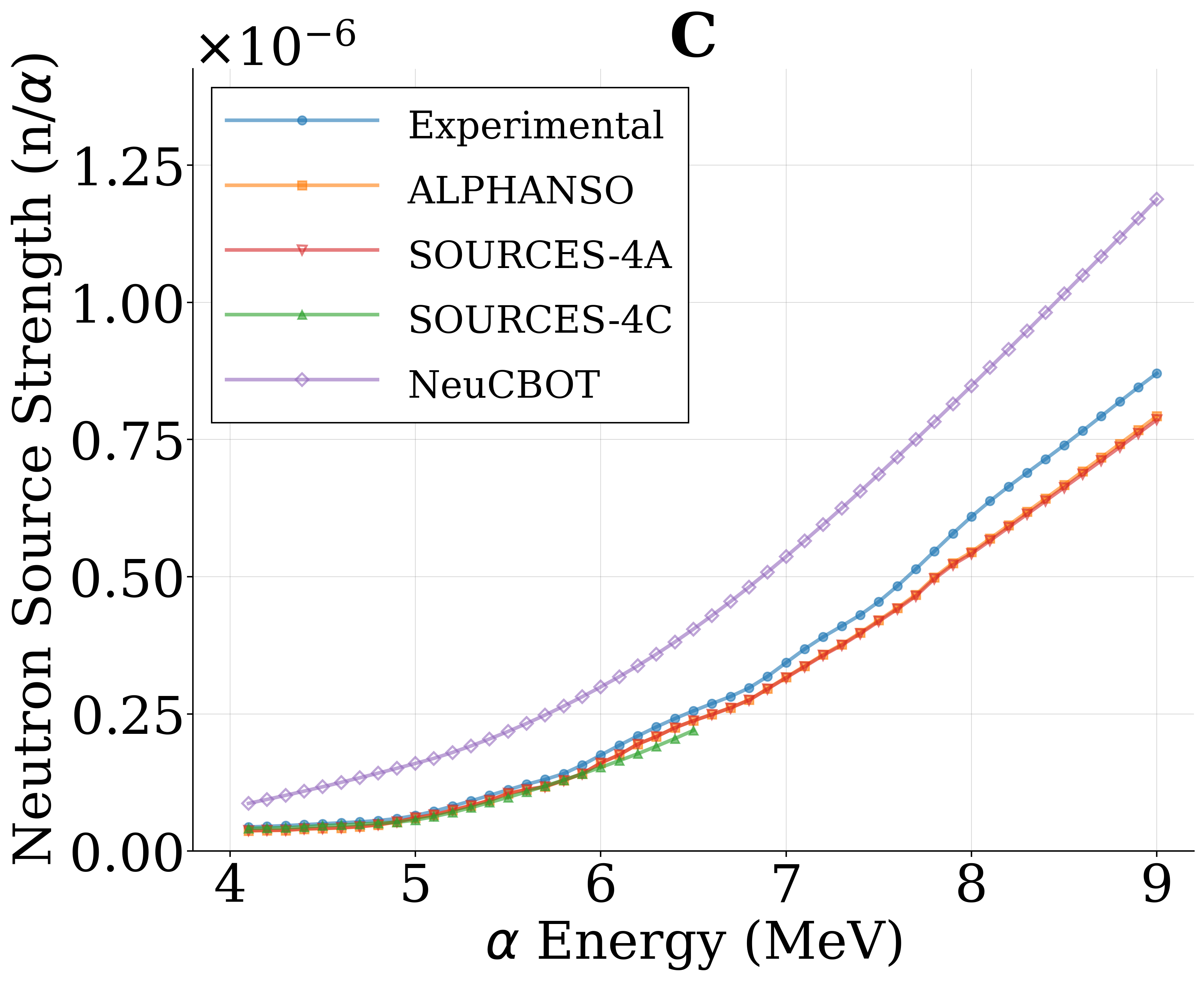}
    \end{minipage}

    \begin{minipage}[t]{0.5\textwidth}
        \centering
        \includegraphics[width=\linewidth]{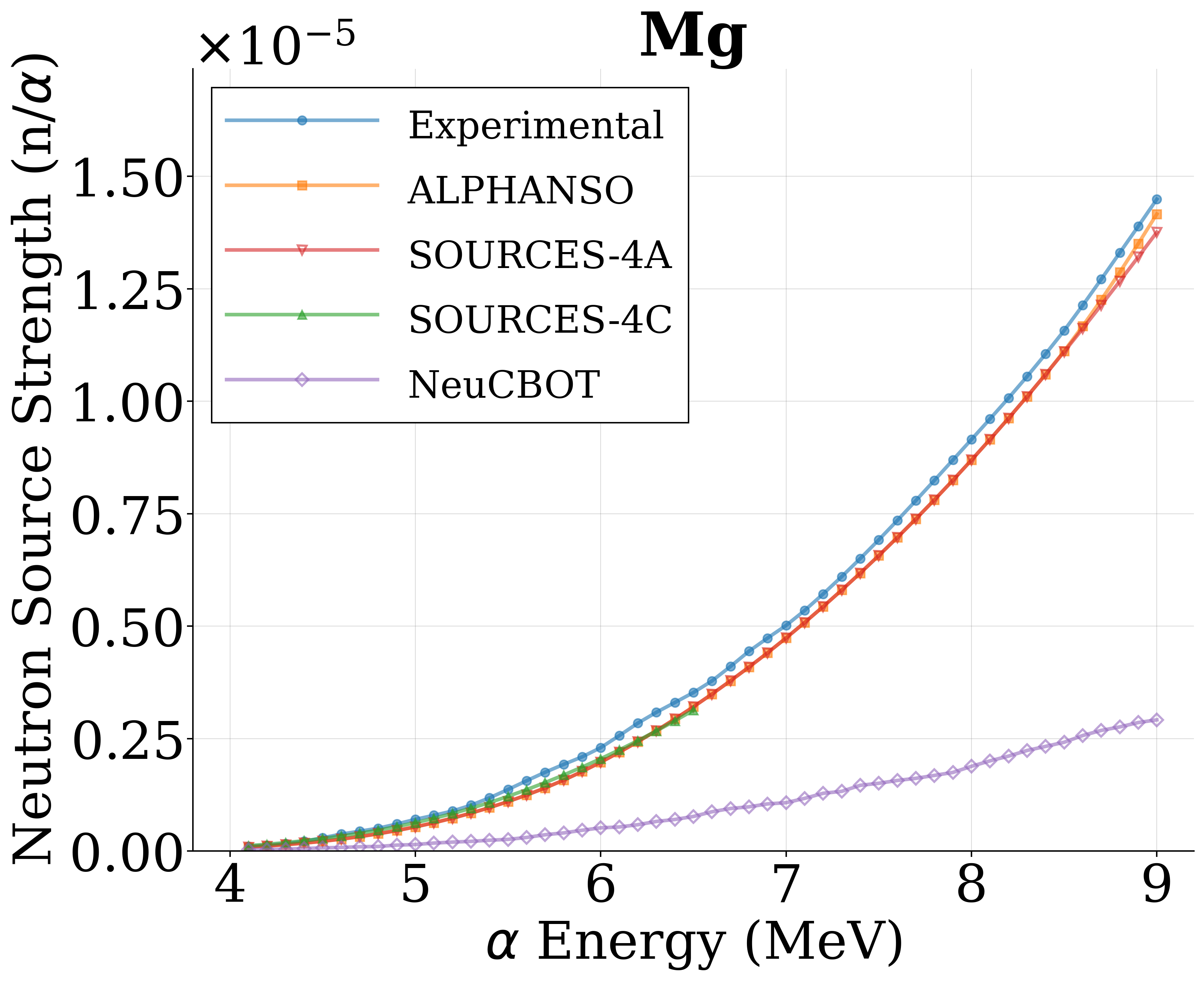}
    \end{minipage}\hfill
    \begin{minipage}[t]{0.5\textwidth}
        \centering
        \includegraphics[width=\linewidth]{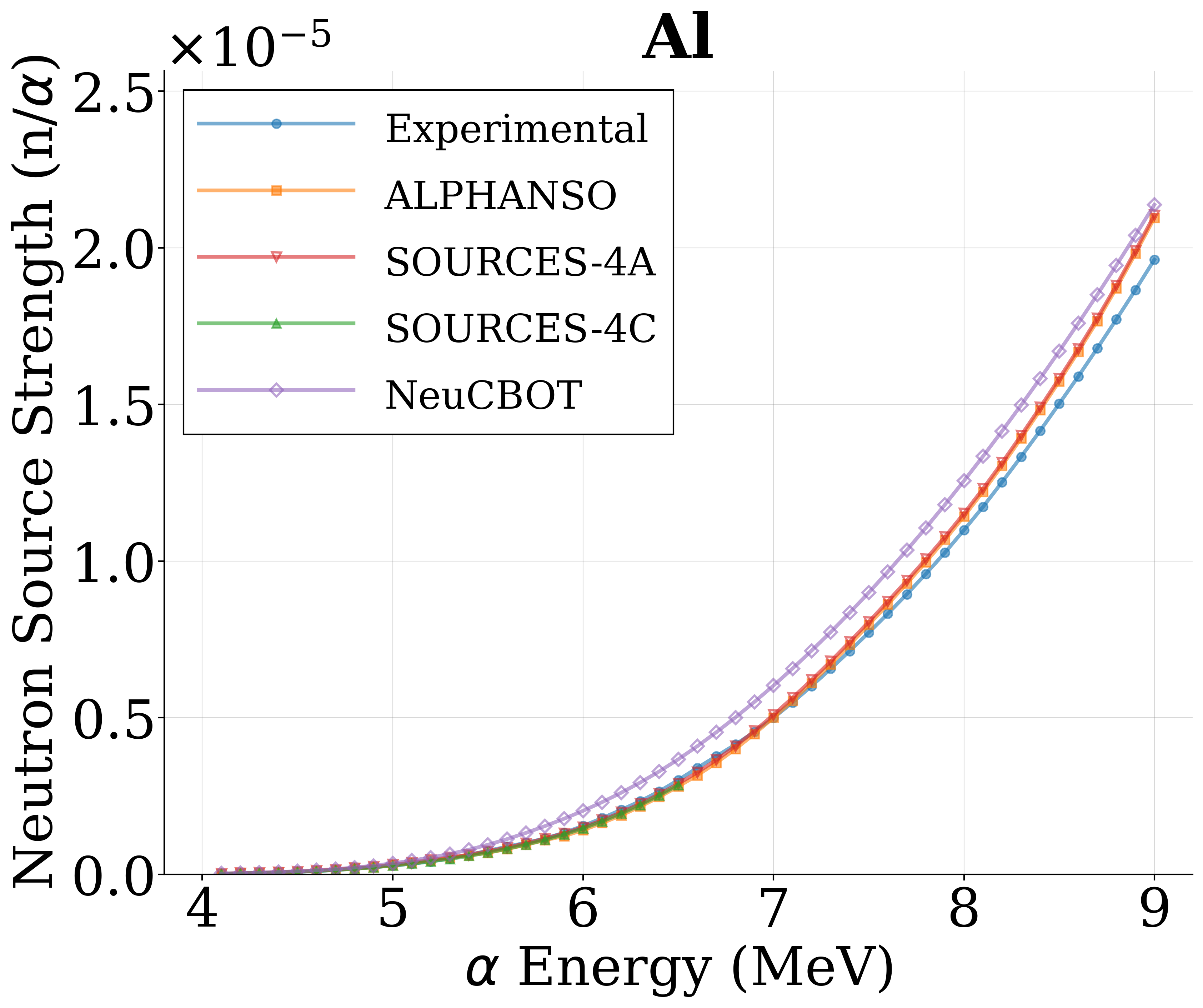}
    \end{minipage}

    \caption{Neutron yields from ALPHANSO, SOURCES-4A, SOURCES-4C, NeuCBOT, and experimental data for $\alpha$-beams incident on selected nuclides. Experimental data taken from \citet{west_measurements_1982}.}
    \label{fig:west}
\end{figure}

We compare the neutron yield computed by ALPHANSO to experimental data from \citet{west_measurements_1982}. These experiments measured the ($\alpha$,n) yield resulting from monoenergetic $\alpha$-beams that span the range of 4 to 9 MeV incident on thick targets. These energies are representative of common particle accelerators and of actinide decay. \Cref{fig:west} gives the computed yields for ALPHANSO, SOURCES-4C, and SOURCES-4A. Recall that SOURCES-4C is limited to $\alpha$-particle energies less than 6.5 MeV. ALPHANSO and SOURCES-4A are in good agreement across the full energy range for all tested materials.

\subsection{Neutron Emission Spectra}

\begin{figure}[htbp]
    \centering

    \begin{minipage}[t]{0.5\textwidth}
        \centering
        \includegraphics[width=\linewidth]{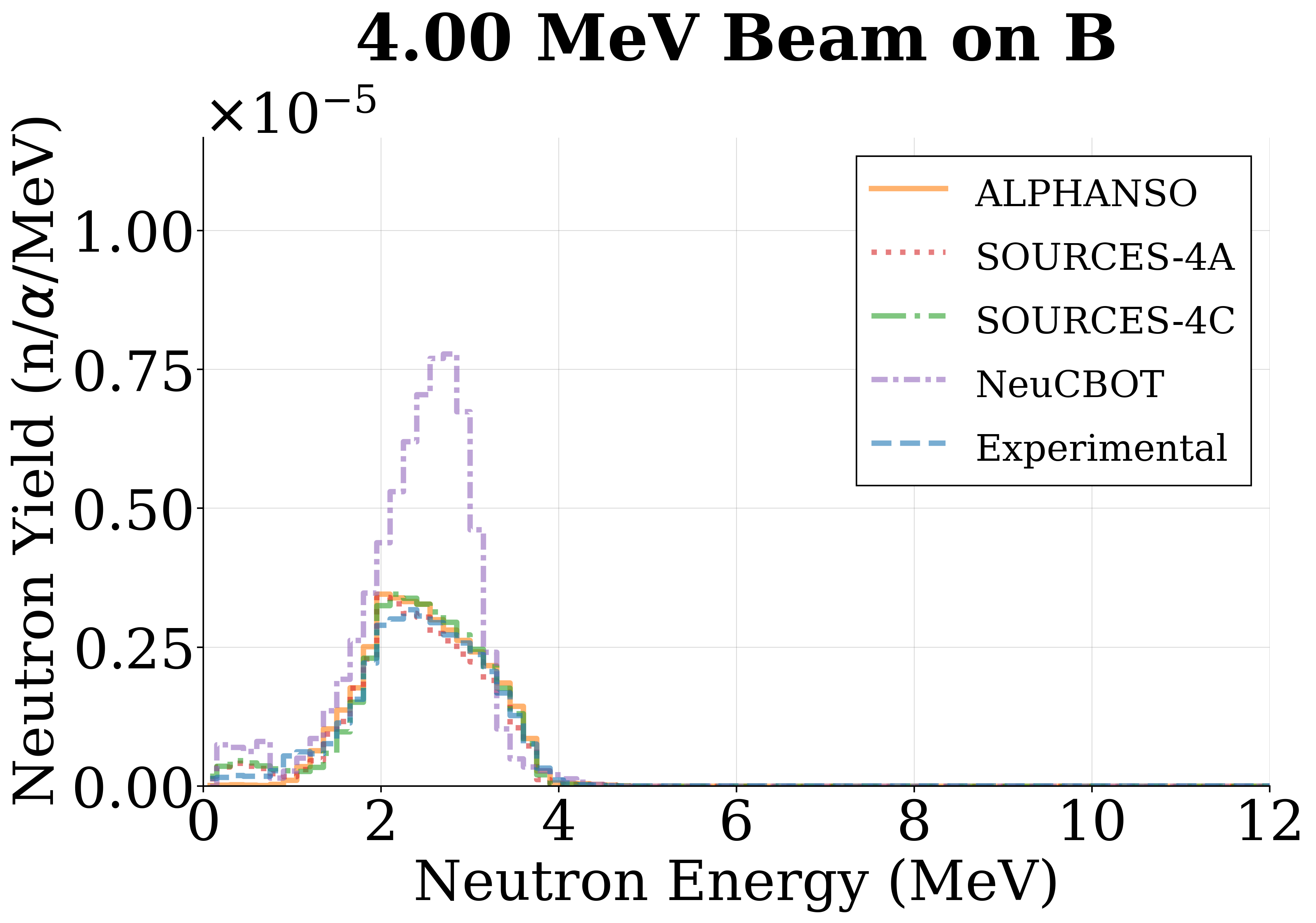}
    \end{minipage}\hfill
    \begin{minipage}[t]{0.5\textwidth}
        \centering
        \includegraphics[width=\linewidth]{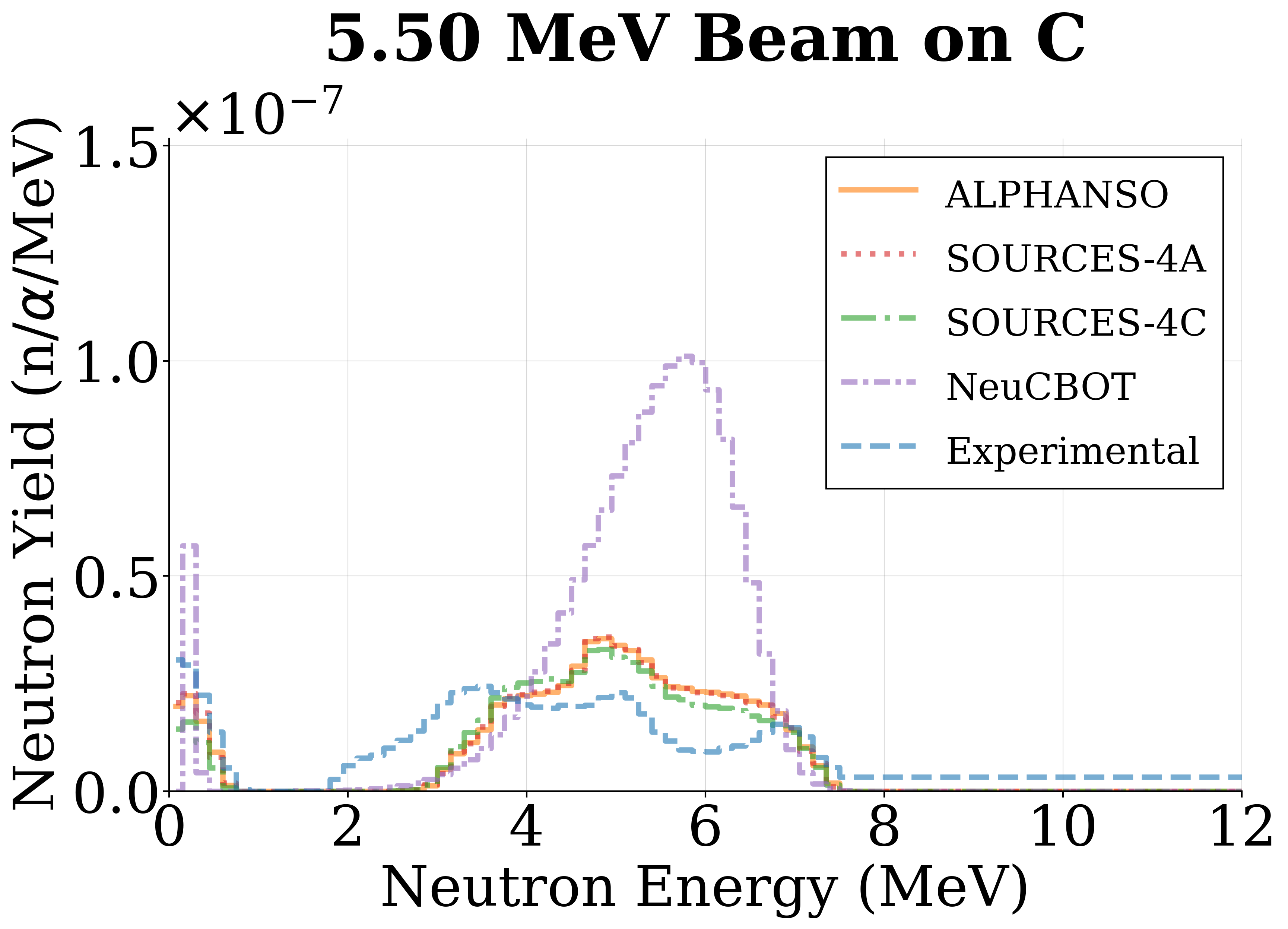}
    \end{minipage}

    \begin{minipage}[t]{0.5\textwidth}
        \centering
        \includegraphics[width=\linewidth]{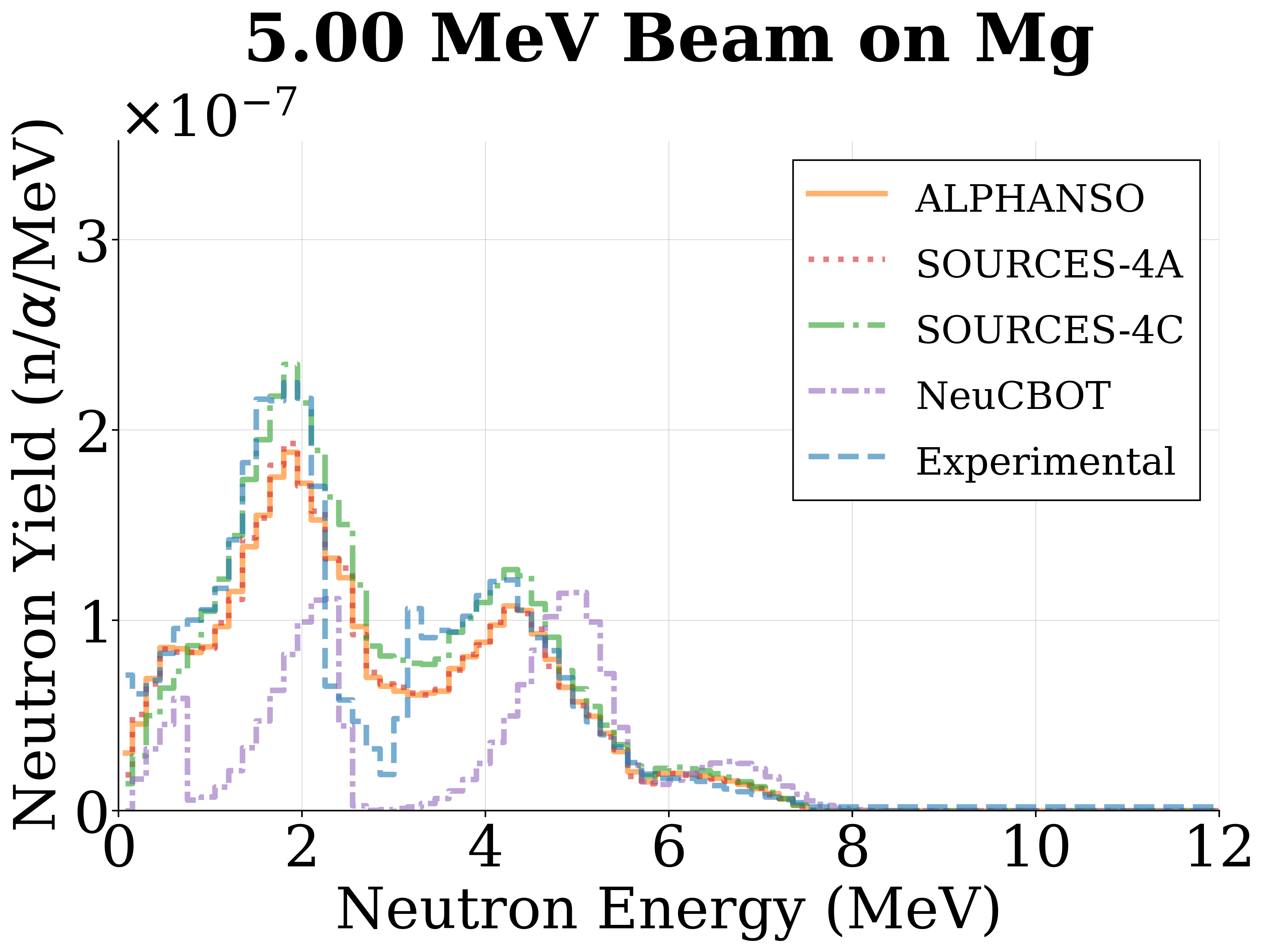}
    \end{minipage}\hfill
    \begin{minipage}[t]{0.5\textwidth}
        \centering
        \includegraphics[width=\linewidth]{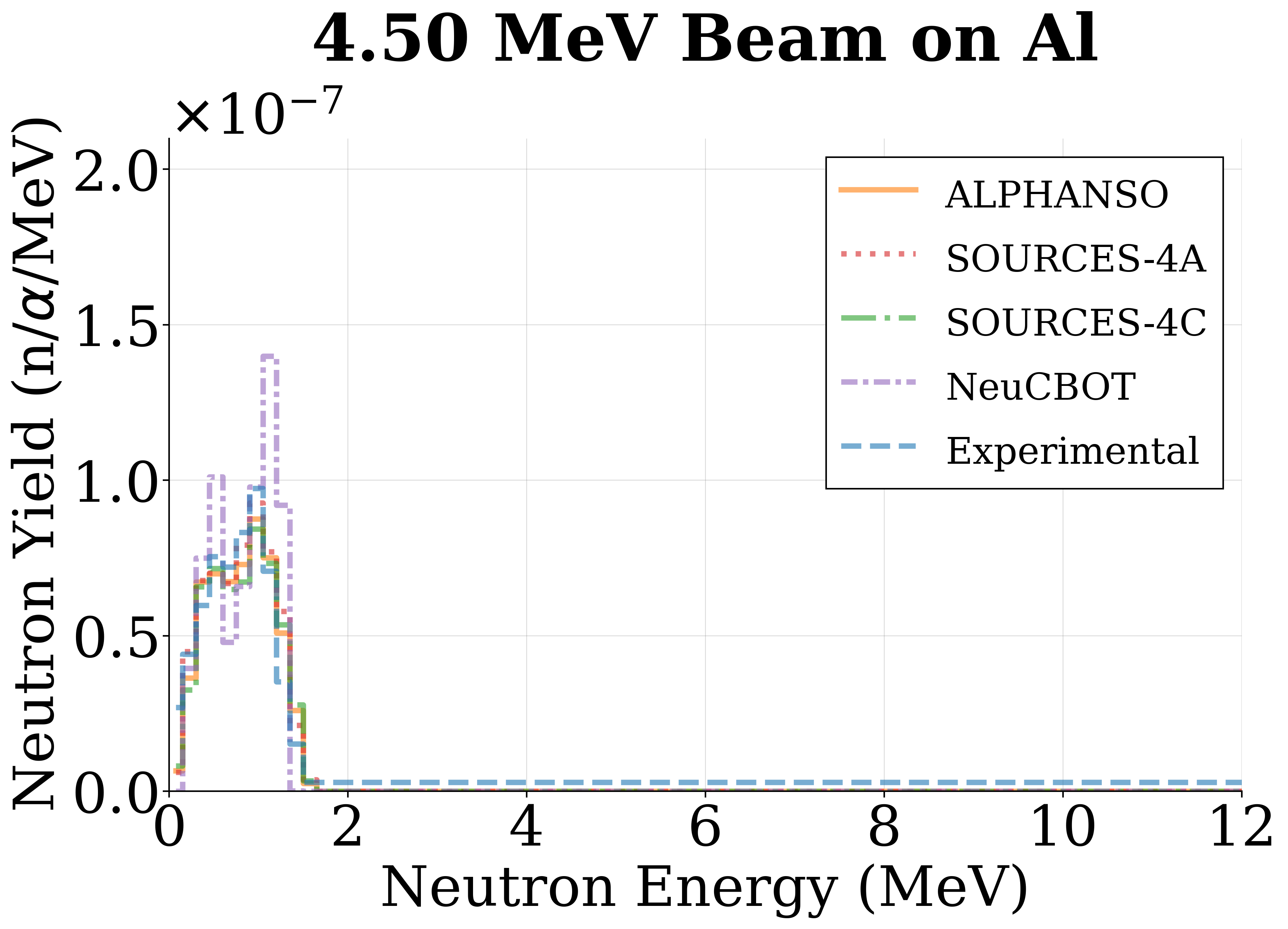}
    \end{minipage}

    \caption{Absolute neutron emission spectra (n/$\alpha$/MeV) from ALPHANSO, SOURCES-4A, and SOURCES-4C, and NeuCBOT, with experimental data for selected nuclides. Experimental data from \citet{jacobs_energy_1983}.}
    \label{fig:spectra}
\end{figure}

\Cref{fig:spectra} shows the absolute neutron emission spectra computed by each code for mono-energetic $\alpha$-particle beams incident on 4 selected thick target elements, compared to experimental data from \citet{jacobs_energy_1983}. We note that using TENDL-2023 data may give inaccurate results, since the ($\alpha$,n) cross sections given in TENDL differ significantly from the evaluated values given in JENDL-5 (see \cref{fig:an_xs}). ALPHANSO shows good agreement with both SOURCES-4A and experimental data across all tested elements.

\subsection{Homogeneous Mixtures}

\begin{figure}[htbp]
    \centering
    \begin{minipage}[t]{0.6\textwidth}
        \centering
        \includegraphics[width=\linewidth]{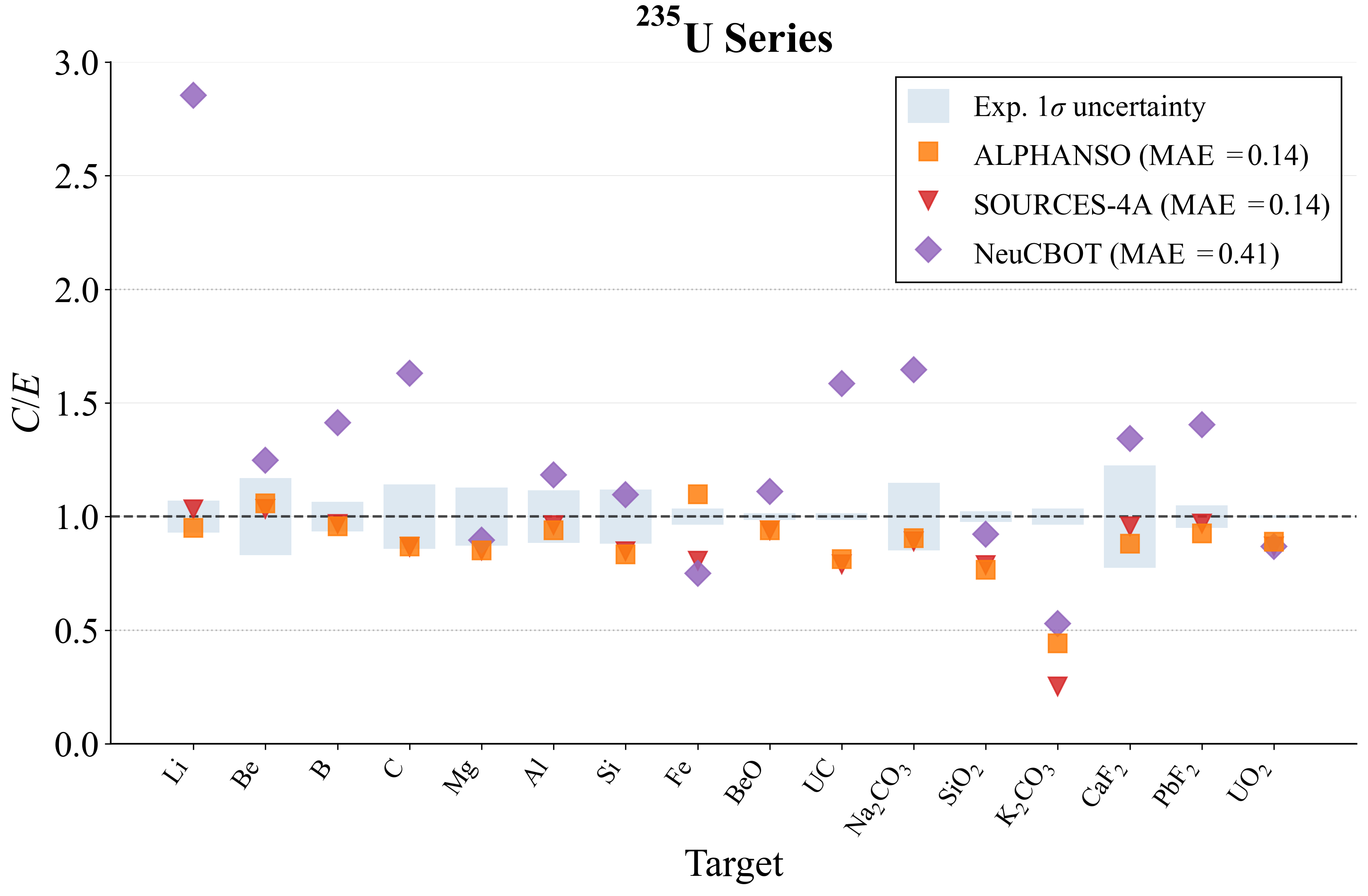}
    \end{minipage}
    \begin{minipage}[t]{0.6\textwidth}
        \centering
        \includegraphics[width=\linewidth]{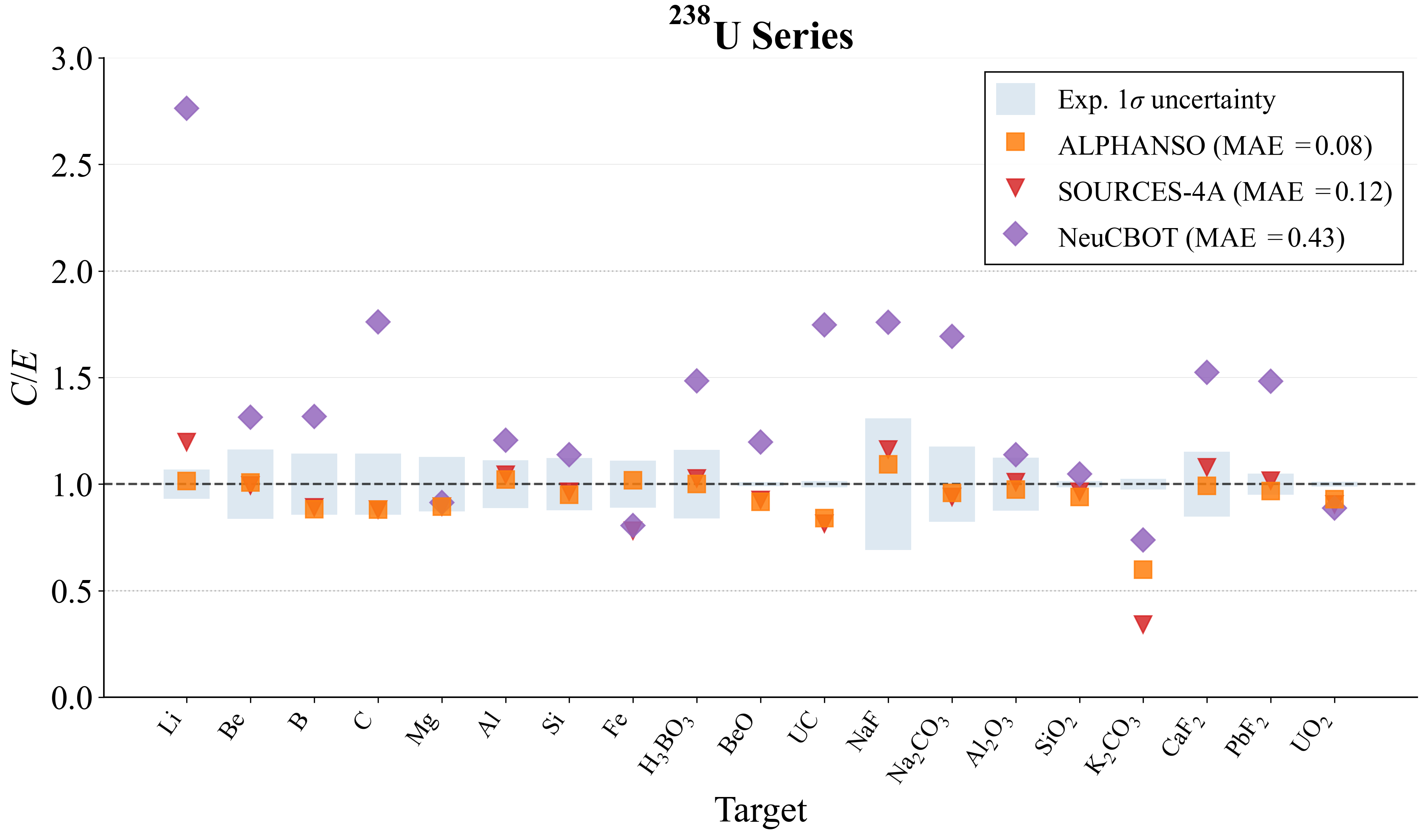}
    \end{minipage}
    \begin{minipage}[t]{0.6\textwidth}
        \centering
        \includegraphics[width=\linewidth]{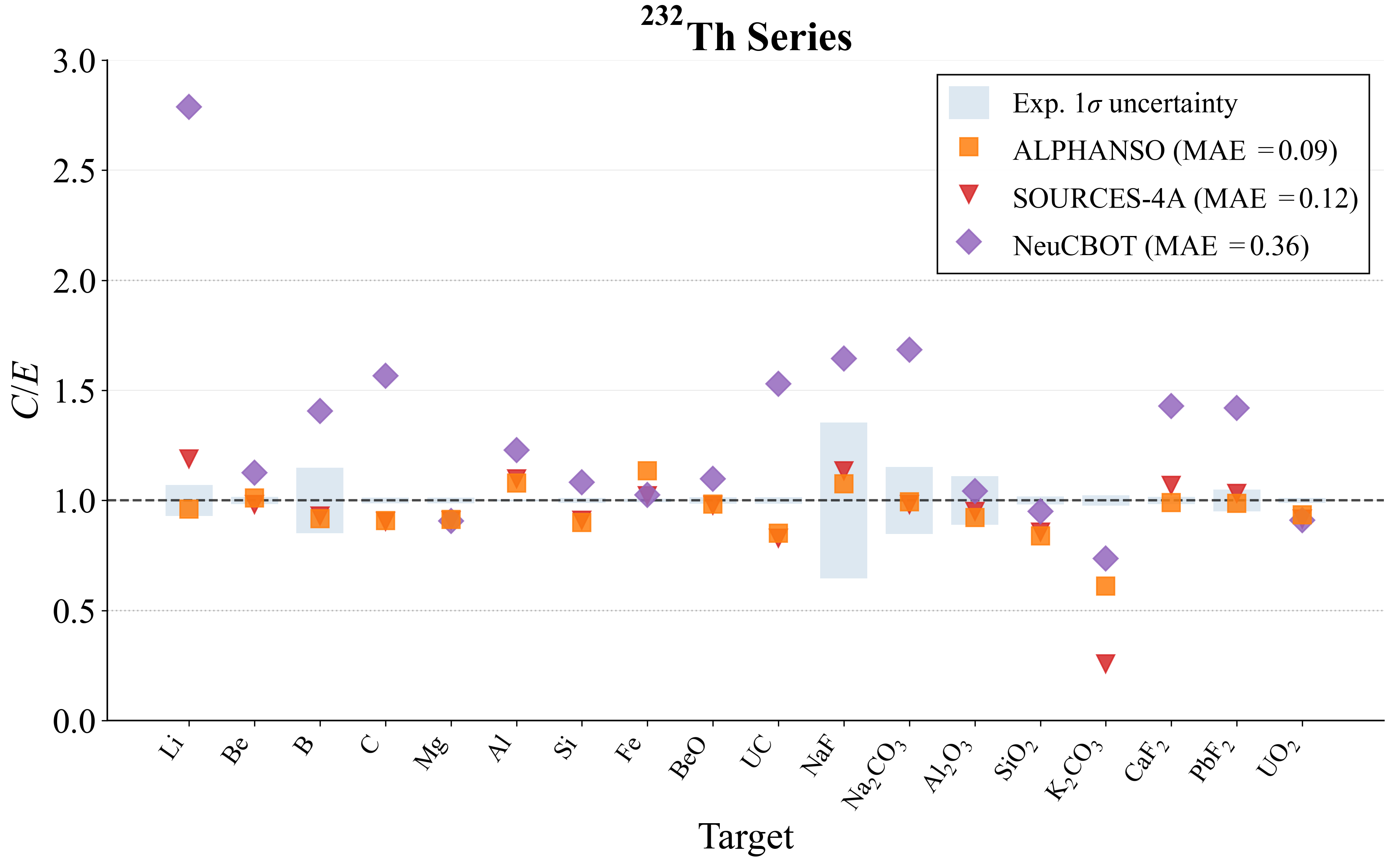}
    \end{minipage}

    \caption{Calculated-to-Experimental (C/E) neutron yield ratios for ($\alpha$,n) reactions on light elements and compounds, for the $^{235}$U, $^{238}$U, and $^{232}$Th $\alpha$ decay series in secular equilibrium. Shaded bars show the $1\sigma$ experimental uncertainty on each measurement. Experimental yields are taken from \citet{mendoza_neutron_2020} for elemental targets and from \citet{fernandes_comparison_2017} for compounds, where the latter compiled data from \citet{west_measurements_1982}, \citet{https://doi.org/10.1029/JB073i010p03135}, and other sources. MAE is computed over all targets for ALPHANSO and SOURCES-4A; SOURCES-4C is omitted as its 6.5~MeV database limit causes systematic underprediction for decay series containing higher-energy $\alpha$ lines.}
    
    \label{fig:u_pu}
\end{figure}

\Cref{fig:u_pu} presents the calculated neutron yield for the emission spectra of $^{232}$Th, $^{235}$U, and $^{238}$U $\alpha$ decay series in secular equilibrium. Where ALPHANSO and SOURCES-4A use the same ($\alpha$,n) cross-section library for a given target nuclide (see Table~\ref{tab:data-sources}), close agreement between the two codes is expected, with residual differences attributable to stopping power and decay spectrum construction; larger deviations indicate a difference in the underlying cross-section data. We compute the Calculated/Experimental (C/E) for ALPHANSO, SOURCES-4A, and NeuCBOT (SOURCES-4C is omitted as it is limited to $E_\alpha \leq 6.5$ MeV and yields systematic underprediction). ALPHANSO is in good agreement with both experimental data and SOURCES-4A. The exception is K$_2$CO$_3$, for which both ALPHANSO and SOURCES-4A substantially underpredict the reference yield; we attribute this to the limited reliability of the underlying \citet{https://doi.org/10.1029/JB073i010p03135} beam data for potassium targets, as compiled by \citet{fernandes_comparison_2017}. NeuCBOT shows larger deviations for several materials, which we attribute to its exclusive use of TALYS-based cross sections.

\subsection{Impact of Data Sources}\label{subsec:data_source_analysis}

\begin{figure}[!h]
    \centering

    \begin{minipage}[t]{0.5\textwidth}
        \centering
        \includegraphics[width=\linewidth]{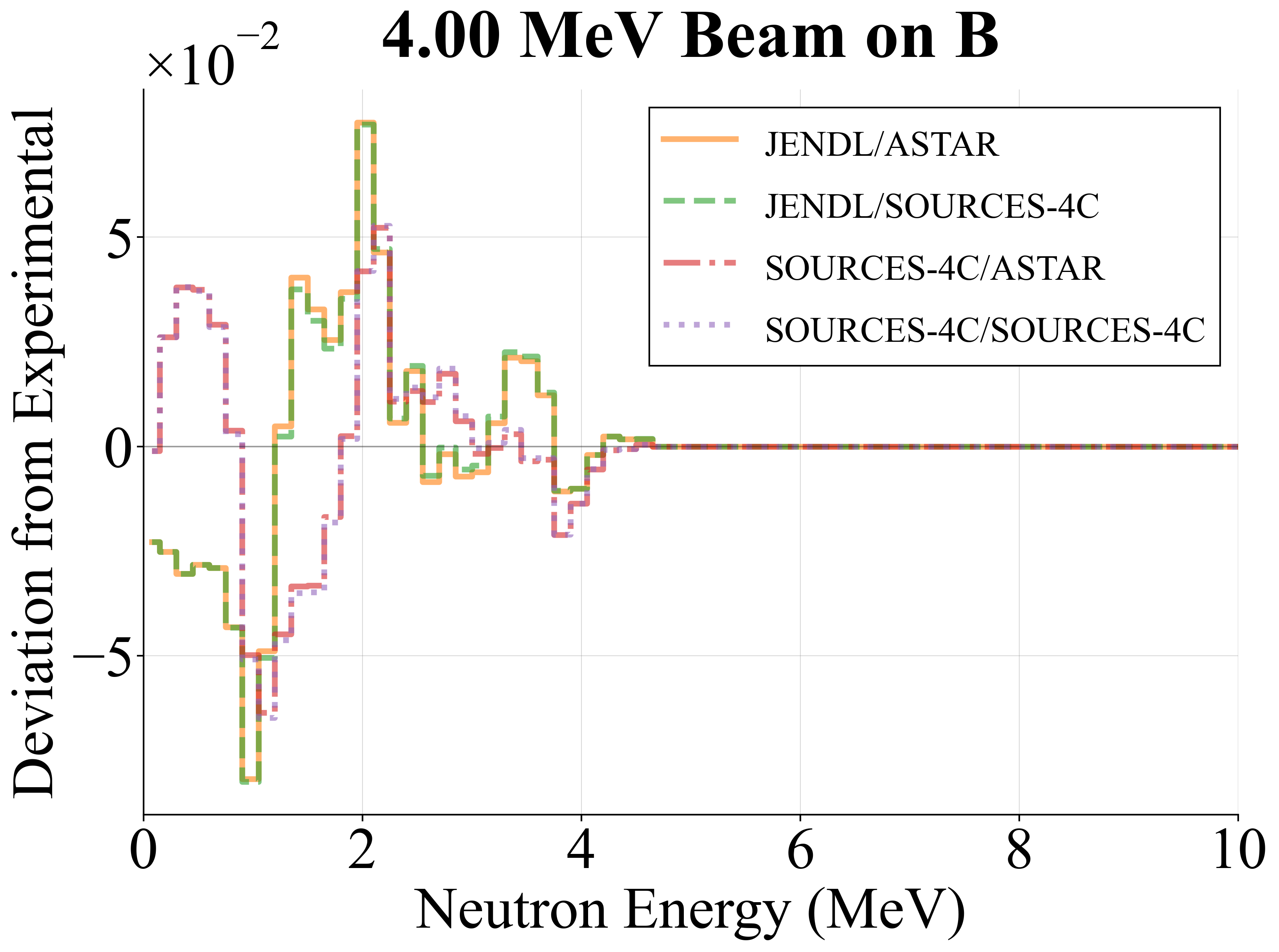}
    \end{minipage}\hfill
    \begin{minipage}[t]{0.5\textwidth}
        \centering
        \includegraphics[width=\linewidth]{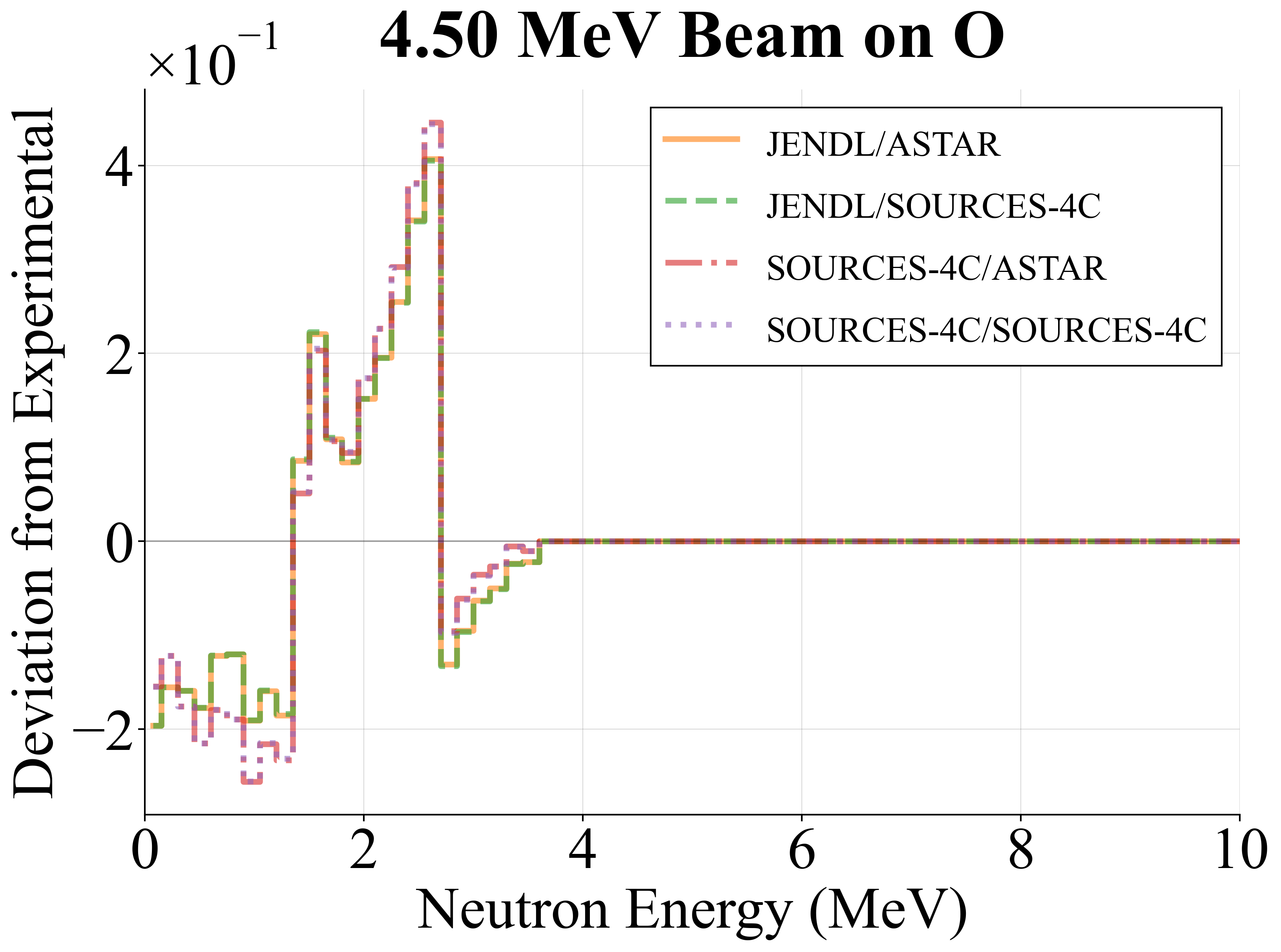}
    \end{minipage}

    \begin{minipage}[t]{0.5\textwidth}
        \centering
        \includegraphics[width=\linewidth]{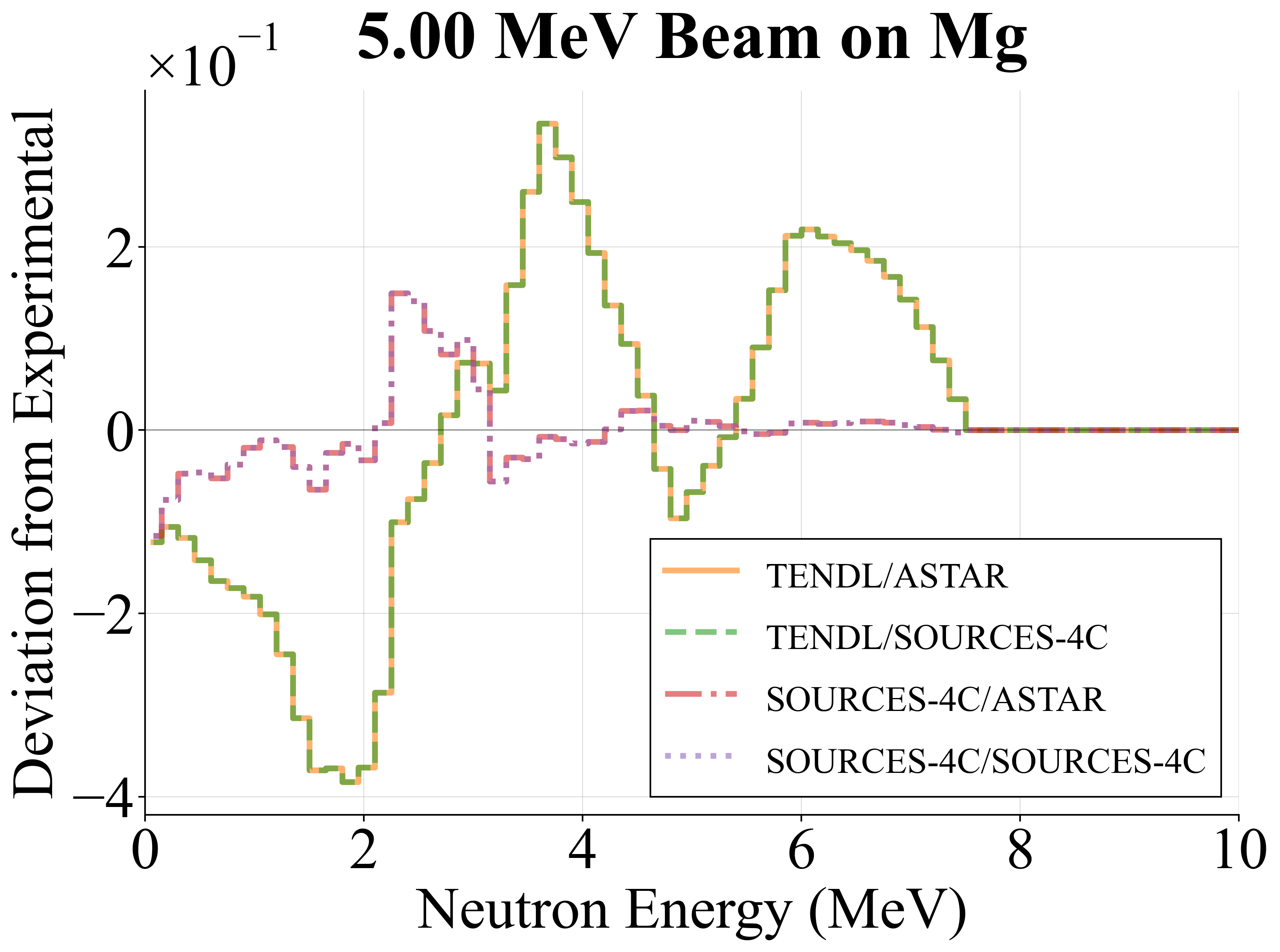}
    \end{minipage}\hfill
    \begin{minipage}[t]{0.5\textwidth}
        \centering
        \includegraphics[width=\linewidth]{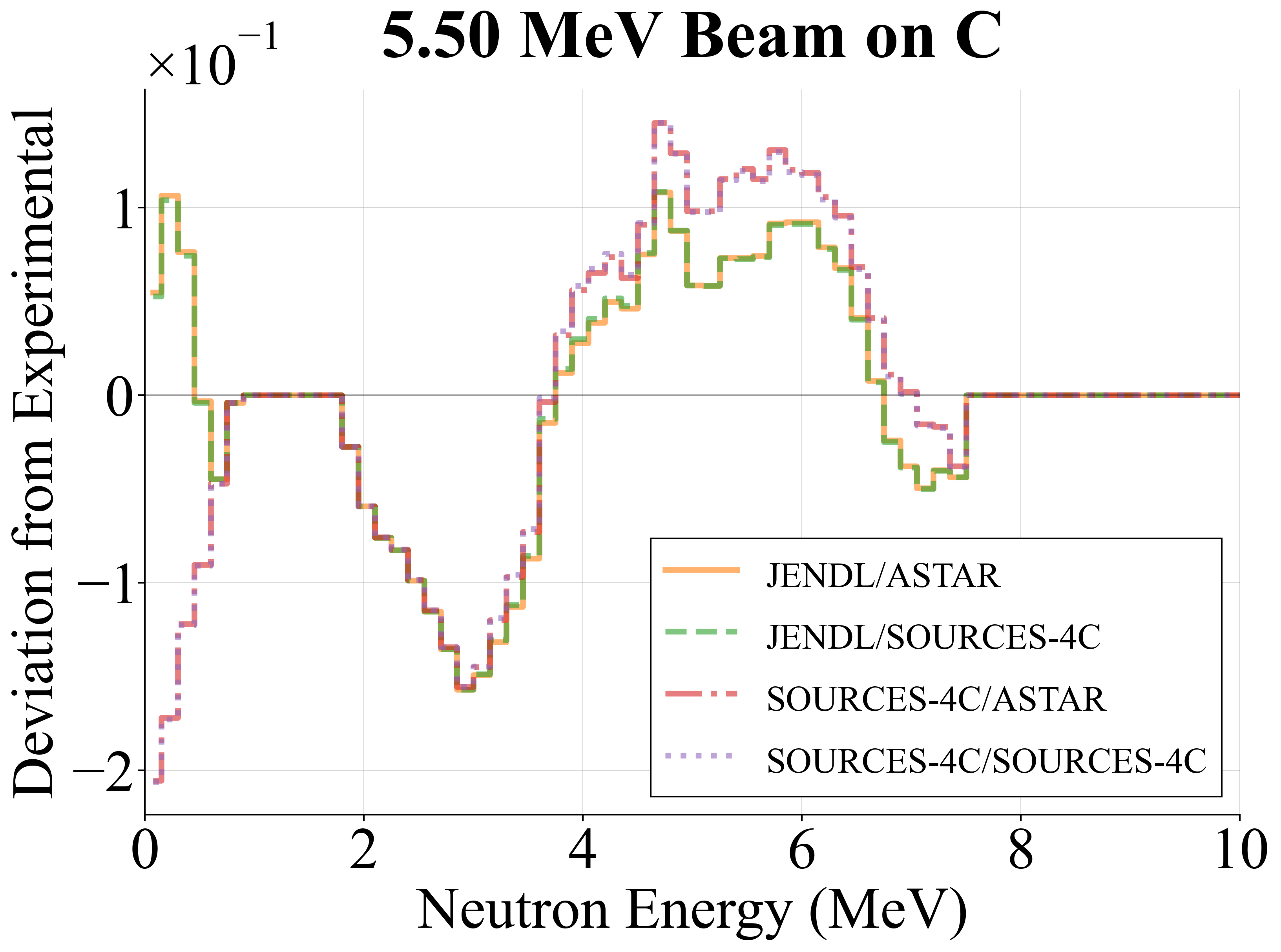}
    \end{minipage}

    \caption{Normalized spectra deviations from experimental data for ALPHANSO computed with different data sources. Legend is in the format ($\alpha$,n) data / stopping power data. The SOURCES-4C label denotes default SOURCES-4C data.}
    \label{fig:data_source_analysis}
\end{figure}

\Cref{fig:data_source_analysis} shows normalized neutron emission spectra computed by ALPHANSO using different data sources for ($\alpha$,n) cross sections and stopping power. As evidenced in \cref{fig:stopping}, stopping powers used by all codes are mostly similar, the main difference is ($\alpha$,n) cross sections. This is confirmed in \cref{fig:data_source_analysis}, where spectra and yields are primarily dependent on the ($\alpha$,n) cross section data used. These results show that future investments into nuclear data improvements should be made into the cross sections and not the stopping powers, because the integral parameters predicted by ALPHANSO show little sensitivity to stopping powers.

\section{Conclusion}\label{sec:conc}

Benchmarking against experimental yields and spectra shows that ALPHANSO reproduces experimental results in good agreement with SOURCES-4A and experimental data. The largest discrepancies arise for nuclides for which only TENDL data is available, highlighting the limitations of TALYS-based evaluations for ($\alpha$,n) reactions at low energies.

Across the tested energy range, differences between codes are primarily driven by the underlying nuclear data. Where JENDL data is available, both yields and neutron energy spectra show excellent correlation with measurements, but when ALPHANSO must fall back to TENDL, errors increase. This suggests that the dominant source of error in ($\alpha$,n) calculations now lies in the ($\alpha$,n) cross section data rather than in the modeling framework or stopping-power data. This highlights the broader role of ALPHANSO as a framework that can rapidly incorporate new evaluations when such data become available.

In summary, ALPHANSO provides a robust, modular, and transparent framework for ($\alpha$,n) source term calculations. Its Python-based architecture, open-source availability, and use of modern GNDS-formatted nuclear data make it straightforward to extend and integrate into existing Python simulation workflows. The code shows good agreement with SOURCES-4A while offering greater flexibility, maintainability, and data transparency: it directly ingests GNDS-formatted evaluated libraries without any reformatting step, and new evaluations can be substituted immediately. ALPHANSO is therefore positioned as a practical and fully validated replacement for legacy ($\alpha$,n) tools such as SOURCES-4C in both research and applied nuclear modeling environments.


\section{Acknowledgments and Funding}

This material is based upon work supported in part by the Department of Energy National Nuclear Security Administration through the Nuclear Science and Security Consortium under Award Number DE-NA0003996. This work was also performed under the auspices of the U.S. Department of Energy by Lawrence Livermore National Laboratory under Contract DE-AC52-07NA27344.

\bibliographystyle{elsarticle-num-names}

\bibliography{references}

@techreport{wilson_sources4c_2002,
  title = {SOURCES 4C: A Code for Calculating ($\alpha$,n), Spontaneous Fission, and Delayed Neutron Sources and Spectra},
  author = {Wilson, W. B. and Perry, R. T. and Shores, E. F. and Charlton, W. S. and Parish, T. A. and Estes, G. P. and Brown, T. H. and Arthur, E. D. and Bozoian, M.},
  institution = {Los Alamos National Laboratory},
  address = {Los Alamos, NM},
  number = {LA-UR-02-1839},
  year = {2002},
  type = {Technical Report},
  language = {en}
}

@article{wrro238795,
          number = {2},
            year = {2026},
           title = {Review of neutron yield from ({\ensuremath{\alpha}}, n) reactions: data, methods, and prospects},
       publisher = {IOP Publishing},
         journal = {Journal of Physics G: Nuclear and Particle Physics},
            note = {{\copyright} 2026 The Author(s). Published by IOP Publishing Ltd. Original content from this work may be used under the terms of the Creative Commons Attribution 4.0 licence (https://creativecommons.org/licenses/by/4.0/). Any further distribution of this work must maintain attribution to the author(s) and the title of the work, journal citation and DOI.},
           month = {February},
             doi = {10.1088/1361-6471/adeffa},
          volume = {53},
             url = {https://eprints.whiterose.ac.uk/id/eprint/238795/},
            issn = {0954-3899},
          author = {Cano-Ott, D. and Cebri{\'a}n, S. and Dimitriou, P. and Gromov, M. and Hara{\'n}czyk, M. and Kish, A. and Kluck, H. and Kudryavtsev, V. A. and Lazanu, I. and Lozza, V. and Luz{\'o}n, G. and Mendoza, E. and Parvu, M. and Pesudo, V. and Pocar, A. and Santorelli, R. and Selvi, M. and Westerdale, S. and Zuzel, G.}
}

@article{https://doi.org/10.1029/JB073i010p03135,
author = {Feige, Y. and Oltman, B. G. and Kastner, J.},
title = {Production rates of neutrons in soils due to natural radioactivity},
journal = {Journal of Geophysical Research (1896-1977)},
volume = {73},
number = {10},
pages = {3135-3142},
doi = {https://doi.org/10.1029/JB073i010p03135},
url = {https://agupubs.onlinelibrary.wiley.com/doi/abs/10.1029/JB073i010p03135},
eprint = {https://agupubs.onlinelibrary.wiley.com/doi/pdf/10.1029/JB073i010p03135},
year = {1968}
}

@article{PhysRevD.98.102006,
  title = {DarkSide-50 532-day dark matter search with low-radioactivity argon},
  author = {Agnes, P. and Albuquerque, I. F. M. and Alexander, T. and Alton, A. K. and Araujo, G. R. and Ave, M. and Back, H. O. and Baldin, B. and Batignani, G. and Biery, K. and Bocci, V. and Bonfini, G. and Bonivento, W. and Bottino, B. and Budano, F. and Bussino, S. and Cadeddu, M. and Cadoni, M. and Calaprice, F. and Caminata, A. and Canci, N. and Candela, A. and Caravati, M. and Cariello, M. and Carlini, M. and Carpinelli, M. and Catalanotti, S. and Cataudella, V. and Cavalcante, P. and Cavuoti, S. and Chepurnov, A. and Cical\`o, C. and Cocco, A. G. and Covone, G. and D'Angelo, D. and D'Incecco, M. and D'Urso, D. and Davini, S. and De Candia, A. and De Cecco, S. and De Deo, M. and De Filippis, G. and De Rosa, G. and De Vincenzi, M. and Derbin, A. V. and Devoto, A. and Di Eusanio, F. and Di Pietro, G. and Dionisi, C. and Downing, M. and Edkins, E. and Empl, A. and Fan, A. and Fiorillo, G. and Fitzpatrick, R. S. and Fomenko, K. and Franco, D. and Gabriele, F. and Galbiati, C. and Ghiano, C. and Giagu, S. and Giganti, C. and Giovanetti, G. K. and Gorchakov, O. and Goretti, A. M. and Granato, F. and Gromov, M. and Guan, M. and Guardincerri, Y. and Gulino, M. and Hackett, B. R. and Herner, K. and Hosseini, B. and Hughes, D. and Humble, P. and Hungerford, E. V. and Ianni, An. and Ippolito, V. and James, I. and Johnson, T. N. and Keeter, K. and Kendziora, C. L. and Kochanek, I. and Koh, G. and Korablev, D. and Korga, G. and Kubankin, A. and Kuss, M. and La Commara, M. and Lai, M. and Li, X. and Lissia, M. and Longo, G. and Ma, Y. and Machado, A. A. and Machulin, I. N. and Mandarano, A. and Mapelli, L. and Mari, S. M. and Maricic, J. and Martoff, C. J. and Messina, A. and Meyers, P. D. and Milincic, R. and Monte, A. and Morrocchi, M. and Mount, B. J. and Muratova, V. N. and Musico, P. and Navrer Agasson, A. and Nozdrina, A. O. and Oleinik, A. and Orsini, M. and Ortica, F. and Pagani, L. and Pallavicini, M. and Pandola, L. and Pantic, E. and Paoloni, E. and Pelczar, K. and Pelliccia, N. and Pocar, A. and Pordes, S. and Poudel, S. S. and Pugachev, D. A. and Qian, H. and Ragusa, F. and Razeti, M. and Razeto, A. and Reinhold, B. and Renshaw, A. L. and Rescigno, M. and Riffard, Q. and Romani, A. and Rossi, B. and Rossi, N. and Sablone, D. and Samoylov, O. and Sands, W. and Sanfilippo, S. and Savarese, C. and Schlitzer, B. and Segreto, E. and Semenov, D. A. and Shchagin, A. and Sheshukov, A. and Singh, P. N. and Skorokhvatov, M. D. and Smirnov, O. and Sotnikov, A. and Stanford, C. and Stracka, S. and Suvorov, Y. and Tartaglia, R. and Testera, G. and Tonazzo, A. and Trinchese, P. and Unzhakov, E. V. and Verducci, M. and Vishneva, A. and Vogelaar, B. and Wada, M. and Waldrop, T. J. and Wang, H. and Wang, Y. and Watson, A. W. and Westerdale, S. and Wojcik, M. M. and Xiang, X. and Xiao, X. and Yang, C. and Ye, Z. and Zhu, C. and Zuzel, G.},
  collaboration = {DarkSide Collaboration},
  journal = {Phys. Rev. D},
  volume = {98},
  issue = {10},
  pages = {102006},
  numpages = {17},
  year = {2018},
  month = {Nov},
  publisher = {American Physical Society},
  doi = {10.1103/PhysRevD.98.102006},
  url = {https://link.aps.org/doi/10.1103/PhysRevD.98.102006}
}

@misc{osti_2571012,
  author       = {Nobre, G. and Capote, R. and Pigni, M. and Trkov, A. and Mattoon, C. and Neudecker, D. and Brown, D. and Chadwick, M. and Kahler, A. and Kleedtke, N. and others},
  title        = {ENDF/B-VIII.1: Alpha Reaction Sublibrary},
  doi          = {10.11578/endf/2571012},
  url          = {https://www.osti.gov/biblio/2571012},
  place        = {United States},
  year         = {2024},
  month        = {08}}

@techreport{osti_1771892,
  author       = {Romano, Catherine E. and Brown, David A. and Croft, Stephen and Favalli, Andrea and Nakae, Les and Pigni, Marco T. and Smith, Michael Scott and Skutnik, Steve and Wieselquist, William and Zerkle, Michael},
  title        = {($\alpha$,n) nuclear data scoping study},
  institution  = {Oak Ridge National Laboratory (ORNL), Oak Ridge, TN (United States)},
  doi          = {10.2172/1771892},
  url          = {https://www.osti.gov/biblio/1771892},
  place        = {United States},
  year         = {2020},
  month        = {10}}

@article{HERMAN20072655,
title = {EMPIRE: Nuclear Reaction Model Code System for Data Evaluation},
journal = {Nuclear Data Sheets},
volume = {108},
number = {12},
pages = {2655-2715},
year = {2007},
note = {Special Issue on Evaluations of Neutron Cross Sections},
issn = {0090-3752},
doi = {https://doi.org/10.1016/j.nds.2007.11.003},
url = {https://www.sciencedirect.com/science/article/pii/S0090375207000981},
author = {M. Herman and R. Capote and B.V. Carlson and P. Obložinský and M. Sin and A. Trkov and H. Wienke and V. Zerkin}
}

@article{GRIESHEIMER20171199,
title = {In-line ($\alpha$,n) source sampling methodology for monte carlo radiation transport simulations},
journal = {Nuclear Engineering and Technology},
volume = {49},
number = {6},
pages = {1199-1210},
year = {2017},
note = {Special Issue on International Conference on Mathematics and Computational Methods Applied to Nuclear Science and Engineering 2017 (M\&C 2017)},
issn = {1738-5733},
doi = {https://doi.org/10.1016/j.net.2017.08.004},
url = {https://www.sciencedirect.com/science/article/pii/S1738573317303017},
author = {David P. Griesheimer and Andrew T. Pavlou and Jason T. Thompson and Jesse C. Holmes and Michael L. Zerkle and Edmund Caro and Hansem Joo}
}

@article{Parvu2025Optimised,
  title        = {Optimised neutron yield calculations from (\(\alpha\),n) reactions with the modified {SOURCES}4 code},
  author       = {Parvu, M. and Krawczun, P. and Kudryavtsev, V. A.},
  journal      = {Applied Radiation and Isotopes},
  volume       = {225},
  pages        = {112035},
  year         = {2025},
  doi          = {10.1016/j.apradiso.2025.112035},
  url          = {https://doi.org/10.1016/j.apradiso.2025.112035}
}

@conference{osti_23178677,
  author       = {Gert, G. and Descalle, M. A. and Mattoon, C. M. and Beck, B. R.},
  title        = {{The LLNL nuclear data infrastructure for the GNDS data format}},
  url          = {https://www.osti.gov/biblio/23178677},
  place        = {United States},
  organization = {American Nuclear Society - ANS, La Grange Park, IL 60526 (United States)},
  publisher    = {ANS - American Nuclear Society; La Grange Park (United States)},
  year         = {2022},
  month        = {07}}

@article{SIMAKOV2017190,
title = {Update of the $\alpha$-n Yields for Reactor Fuel Materials for the Interest of Nuclear Safeguards},
journal = {Nuclear Data Sheets},
volume = {139},
pages = {190-203},
year = {2017},
note = {Special Issue on Nuclear Reaction Data},
issn = {0090-3752},
doi = {https://doi.org/10.1016/j.nds.2017.01.005},
url = {https://www.sciencedirect.com/science/article/pii/S0090375217300054},
author = {S.P. Simakov and Q.Y. {van den Berg}}
}

@conference{osti_6893289,
  author       = {Arthur, E D},
  title        = {The {GNASH} preequilibrium-statistical nuclear model code},
  url          = {https://www.osti.gov/biblio/6893289},
  place        = {United States},
  organization = {Los Alamos National Lab., NM (USA)},
  year         = {1987},
  month        = {12}}

@misc{nobre2025endfbviii1updatednuclearreaction,
      title={ENDF/B-VIII.1: Updated Nuclear Reaction Data Library for Science and Applications}, 
      author={G. P. A. Nobre and R. Capote and M. T. Pigni and A. Trkov and C. M. Mattoon and D. Neudecker and D. A. Brown and M. B. Chadwick and A. C. Kahler and N. A. Kleedtke and M. Zerkle and A. I. Hawari and C. W. Chapman and N. C. Fleming and J. L. Wormald and K. Ramić and Y. Danon and N. A. Gibson and P. Brain and M. W. Paris and G. M. Hale and I. J. Thompson and D. P. Barry and I. Stetcu and W. Haeck and A. E. Lovell and M. R. Mumpower and G. Potel and K. Kravvaris and G. Noguere and J. D. McDonnell and A. D. Carlson and M. Dunn and T. Kawano and D. Wiarda and I. Al-Qasir and G. Arbanas and R. Arcilla and B. Beck and D. Bernard and R. Beyer and J. M. Brown and O. Cabellos and R. J. Casperson and Y. Cheng and E. V. Chimanski and R. Coles and M. Cornock and J. Cotchen and J. P. W. Crozier and D. E. Cullen and A. Daskalakis and M. -A. Descalle and D. D. DiJulio and P. Dimitriou and A. C. Dreyfuss and I. Durán and R. Ferrer and T. Gaines and V. Gillette and G. Gert and K. H. Guber and J. D. Haverkamp and M. W. Herman and J. Holmes and M. Hursin and N. Jisrawi and A. R. Junghans and K. J. Kelly and H. I. Kim and K. S. Kim and A. J. Koning and M. Koštál and B. K. Laramee and A. Lauer-Coles and L. Leal and H. Y. Lee and A. M. Lewis and J. Malec and J. I. Márquez Damián and W. J. Marshall and A. Mattera and G. Muhrer and A. Ney and W. E. Ormand and D. K. Parsons and C. M. Percher and V. G. Pronyaev and A. Qteish and S. Quaglioni and M. Rapp and J. J. Ressler and M. Rising and D. Rochman and P. K. Romano and D. Roubtsov and G. Schnabel and M. Schulc and G. J. Siemers and A. A. Sonzogni and P. Talou and J. Thompson and T. H. Trumbull and S. C. van der Marck and M. Vorabbi and C. Wemple and K. A. Wendt and M. White and R. Q. Wright},
      year={2025},
      eprint={2511.03564},
      archivePrefix={arXiv},
      primaryClass={physics.app-ph},
      url={https://arxiv.org/abs/2511.03564}, 
}

@techreport{COG_manual_202590,
	title = {COG User's Manual Sixth Edition},
	language = {en},
	number = {UCRL-TM-202590},
	urldate = {2025-08-28},
	author = {Wilcox, Thomas and Lent, Edward and Buck, Richard and Lee, Chuck},
	month = mar,
    institution = {Lawrence Livermore National Laboratory},
	year = {2024}
}

@article{agostinelli_geant4simulation_2003,
	title = {Geant4—a simulation toolkit},
	volume = {506},
	issn = {0168-9002},
	url = {https://www.sciencedirect.com/science/article/pii/S0168900203013688},
	doi = {10.1016/S0168-9002(03)01368-8},
	number = {3},
	urldate = {2025-08-28},
	journal = {Nuclear Instruments and Methods in Physics Research Section A: Accelerators, Spectrometers, Detectors and Associated Equipment},
	author = {Agostinelli, S. and Allison, J. and Amako, K. and Apostolakis, J. and Araujo, H. and Arce, P. and Asai, M. and Axen, D. and Banerjee, S. and Barrand, G. and Behner, F. and Bellagamba, L. and Boudreau, J. and Broglia, L. and Brunengo, A. and Burkhardt, H. and Chauvie, S. and Chuma, J. and Chytracek, R. and Cooperman, G. and Cosmo, G. and Degtyarenko, P. and Dell'Acqua, A. and Depaola, G. and Dietrich, D. and Enami, R. and Feliciello, A. and Ferguson, C. and Fesefeldt, H. and Folger, G. and Foppiano, F. and Forti, A. and Garelli, S. and Giani, S. and Giannitrapani, R. and Gibin, D. and Gómez Cadenas, J. J. and González, I. and Gracia Abril, G. and Greeniaus, G. and Greiner, W. and Grichine, V. and Grossheim, A. and Guatelli, S. and Gumplinger, P. and Hamatsu, R. and Hashimoto, K. and Hasui, H. and Heikkinen, A. and Howard, A. and Ivanchenko, V. and Johnson, A. and Jones, F. W. and Kallenbach, J. and Kanaya, N. and Kawabata, M. and Kawabata, Y. and Kawaguti, M. and Kelner, S. and Kent, P. and Kimura, A. and Kodama, T. and Kokoulin, R. and Kossov, M. and Kurashige, H. and Lamanna, E. and Lampén, T. and Lara, V. and Lefebure, V. and Lei, F. and Liendl, M. and Lockman, W. and Longo, F. and Magni, S. and Maire, M. and Medernach, E. and Minamimoto, K. and Mora de Freitas, P. and Morita, Y. and Murakami, K. and Nagamatu, M. and Nartallo, R. and Nieminen, P. and Nishimura, T. and Ohtsubo, K. and Okamura, M. and O'Neale, S. and Oohata, Y. and Paech, K. and Perl, J. and Pfeiffer, A. and Pia, M. G. and Ranjard, F. and Rybin, A. and Sadilov, S. and Di Salvo, E. and Santin, G. and Sasaki, T. and Savvas, N. and Sawada, Y. and Scherer, S. and Sei, S. and Sirotenko, V. and Smith, D. and Starkov, N. and Stoecker, H. and Sulkimo, J. and Takahata, M. and Tanaka, S. and Tcherniaev, E. and Safai Tehrani, E. and Tropeano, M. and Truscott, P. and Uno, H. and Urban, L. and Urban, P. and Verderi, M. and Walkden, A. and Wander, W. and Weber, H. and Wellisch, J. P. and Wenaus, T. and Williams, D. C. and Wright, D. and Yamada, T. and Yoshida, H. and Zschiesche, D.},
	month = jul,
	year = {2003},
	pages = {250--303},
}

@article{tomasello_calculation_2008,
	title = {Calculation of neutron background for underground experiments},
	volume = {595},
	issn = {0168-9002},
	url = {https://www.sciencedirect.com/science/article/pii/S0168900208009984},
	doi = {10.1016/j.nima.2008.07.071},
	number = {2},
	urldate = {2025-08-28},
	journal = {Nuclear Instruments and Methods in Physics Research Section A: Accelerators, Spectrometers, Detectors and Associated Equipment},
	author = {Tomasello, V. and Kudryavtsev, V. A. and Robinson, M.},
	month = oct,
	year = {2008},
	pages = {431--438},
}

@article{mendoza_neutron_2020,
	title = {Neutron production induced by $\alpha$-decay with {Geant4}},
	volume = {960},
	issn = {0168-9002},
	url = {https://www.sciencedirect.com/science/article/pii/S0168900220302333},
	doi = {10.1016/j.nima.2020.163659},
	urldate = {2025-08-28},
	journal = {Nuclear Instruments and Methods in Physics Research Section A: Accelerators, Spectrometers, Detectors and Associated Equipment},
	author = {Mendoza, E. and Cano-Ott, D. and Romojaro, P. and Alcayne, V. and García Abia, P. and Pesudo, V. and Romero, L. and Santorelli, R.},
	month = apr,
	year = {2020},
	pages = {163659},
}

@article{mei_evaluation_2009,
	title = {Evaluation of ($\alpha$, n) induced neutrons as a background for dark matter experiments},
	volume = {606},
	issn = {0168-9002},
	doi = {10.1016/j.nima.2009.04.032},
	number = {3},
	urldate = {2025-08-28},
	journal = {Nuclear Instruments and Methods in Physics Research Section A: Accelerators, Spectrometers, Detectors and Associated Equipment},
	author = {Mei, D. -M. and Zhang, C. and Hime, A.},
	month = jul,
	year = {2009},
	pages = {651--660},
}

@article{westerdale_radiogenic_2017,
	title = {Radiogenic {Neutron} {Yield} {Calculations} for {Low}-{Background} {Experiments}},
	volume = {875},
	issn = {01689002},
	url = {http://arxiv.org/abs/1702.02465},
	doi = {10.1016/j.nima.2017.09.007},
	language = {en},
	urldate = {2025-09-01},
	journal = {Nuclear Instruments and Methods in Physics Research Section A: Accelerators, Spectrometers, Detectors and Associated Equipment},
	author = {Westerdale, S. and Meyers, P. D.},
	month = dec,
	year = {2017},
	note = {arXiv:1702.02465 [physics]},
	pages = {57--64},
}

@article{west_measurements_1982,
	title = {Measurements of thick-target ($\alpha$,n) yields from light elements},
	journal = {Annals of Nuclear Energy},
	volume = {9},
	pages = {551--577},
	language = {en},
	author = {West, D. and Sherwood, A. C.},
	year = {1982},
}

@article{vlaskin_neutron_2015,
	title = {Neutron {Yield} of the {Reaction} ($\alpha$, n) on {Thick} {Targets} {Comprised} of {Light} {Elements}},
	volume = {117},
	issn = {1063-4258, 1573-8205},
	url = {http://link.springer.com/10.1007/s10512-015-9933-5},
	doi = {10.1007/s10512-015-9933-5},
	language = {en},
	number = {5},
	urldate = {2025-09-01},
	journal = {Atomic Energy},
	author = {Vlaskin, G. N. and Khomyakov, Yu. S. and Bulanenko, V. I.},
	month = mar,
	year = {2015},
	pages = {357--365},
}

@article{fernandes_comparison_2017,
	title = {Comparison of thick-target (alpha,n) yield calculation codes},
	volume = {153},
	issn = {2100-014X},
	url = {http://www.epj-conferences.org/10.1051/epjconf/201715307021},
	doi = {10.1051/epjconf/201715307021},
	language = {en},
	urldate = {2025-09-01},
	journal = {EPJ Web of Conferences},
	author = {Fernandes, Ana C. and Kling, Andreas and Vlaskin, Gennadiy N.},
	editor = {Malvagi, F. and Malouch, F. and Diop, C.M’B. and Miss, J. and Trama, J.C.},
	year = {2017},
	pages = {07021},
}

@techreport{TechReport_2024_LANL_LA-UR-24-24602Rev.1_KuleszaAdamsEtAl,
  address = {Los Alamos, NM, USA},
  author = {Kulesza, Joel A. and Adams, Terry R. and Armstrong, Jerawan C. and Bolding, Simon R. and Brown, Forrest B. and Bull, Jeffrey S. and Burke, Timothy P. and Clark, Alexander R. and Forster, III, Robert Arthur and Giron, Jesse F. and Grieve, Avery S. and Josey, Colin J. and Martz, Roger L. and McKinney, Gregg W. and Pearson, Eric J. and Rising, Michael E. and Solomon, Jr., Clell J. and Swaminarayan, Sriram and Trahan, Travis J. and Weaver, Colin A. and Wilson, Stephen C. and Zukaitis, Anthony J.},
  doi = {10.2172/2372634},
  editor = {Joel A. Kulesza},
  institution = {Los Alamos National Laboratory},
  month = {May},
  number = {LA-UR-24-24602, Rev.~1},
  title = {{MCNP\textsuperscript{\textregistered} Code Version 6.3.1 Theory \& User Manual}},
  url = {https://www.osti.gov/biblio/2372634},
  year = {2024}
}

@techreport{bell1973origen,
  title={ORIGEN: the ORNL isotope generation and depletion code},
  author={Bell, MJ},
  year={1973},
  institution={Oak Ridge National Lab.(ORNL), Oak Ridge, TN (United States)}
}

@techreport{bates2024mcnpsmesa,
  title={MCNPs Easy Sources for ($\alpha$, n)(MESA) 1.0: A User’s Guide},
  author={Bates, Cameron Russell},
  year={2024},
  institution={Los Alamos National Laboratory (LANL), Los Alamos, NM (United States)}
}

@article{iwamoto_japanese_2023,
	title = {Japanese evaluated nuclear data library version 5: {JENDL}-5},
	volume = {60},
	issn = {0022-3131, 1881-1248},
	shorttitle = {Japanese evaluated nuclear data library version 5},
	url = {https://www.tandfonline.com/doi/full/10.1080/00223131.2022.2141903},
	doi = {10.1080/00223131.2022.2141903},
	language = {en},
	number = {1},
	urldate = {2025-09-01},
	journal = {Journal of Nuclear Science and Technology},
	author = {Iwamoto, Osamu and Iwamoto, Nobuyuki and Kunieda, Satoshi and Minato, Futoshi and Nakayama, Shinsuke and Abe, Yutaka and Tsubakihara, Kohsuke and Okumura, Shin and Ishizuka, Chikako and Yoshida, Tadashi and Chiba, Satoshi and Otuka, Naohiko and Sublet, Jean-Christophe and Iwamoto, Hiroki and Yamamoto, Kazuyoshi and Nagaya, Yasunobu and Tada, Kenichi and Konno, Chikara and Matsuda, Norihiro and Yokoyama, Kenji and Taninaka, Hiroshi and Oizumi, Akito and Fukushima, Masahiro and Okita, Shoichiro and Chiba, Go and Sato, Satoshi and Ohta, Masayuki and Kwon, Saerom},
	month = jan,
	year = {2023},
	pages = {1--60},
}

@article{koning_tendl_2019,
	series = {Special {Issue} on {Nuclear} {Reaction} {Data}},
	title = {{TENDL}: {Complete} {Nuclear} {Data} {Library} for {Innovative} {Nuclear} {Science} and {Technology}},
	volume = {155},
	issn = {0090-3752},
	shorttitle = {{TENDL}},
	url = {https://www.sciencedirect.com/science/article/pii/S009037521930002X},
	doi = {10.1016/j.nds.2019.01.002},
	urldate = {2025-09-01},
	journal = {Nuclear Data Sheets},
	author = {Koning, A. J. and Rochman, D. and Sublet, J. -Ch. and Dzysiuk, N. and Fleming, M. and van der Marck, S.},
	month = jan,
	year = {2019},
	pages = {1--55},
}

@article{berger_stopping-power_2009,
	title = {Stopping-{Power} \& {Range} {Tables} for {Electrons}, {Protons}, and {Helium} {Ions}},
	url = {https://www.nist.gov/pml/stopping-power-range-tables-electrons-protons-and-helium-ions},
	language = {en},
	urldate = {2025-09-01},
	journal = {NIST},
	author = {Berger, M.J. and Coursey, J.S. and Zucker, M.A. and Chang, J.},
	month = oct,
	year = {2009},
	note = {Last Modified: 2024-10-11T12:05-04:00},
}

@article{ziegler_srim_2010,
	title = {{SRIM} - {The} stopping and range of ions in matter (2010)},
	volume = {268},
	issn = {0168-583X},
	url = {https://ui.adsabs.harvard.edu/abs/2010NIMPB.268.1818Z},
	doi = {10.1016/j.nimb.2010.02.091},
	urldate = {2025-09-01},
	journal = {Nuclear Instruments and Methods in Physics Research B},
	author = {Ziegler, James F. and Ziegler, M. D. and Biersack, J. P.},
	month = jun,
	year = {2010},
	note = {ADS Bibcode: 2010NIMPB.268.1818Z},
	pages = {1818--1823},
}

@book{ziegler_helium_1977,
	title = {Helium: {Stopping} {Powers} and {Ranges} in {All} {Elemental} {Matter}},
	isbn = {978-0-08-021606-5},
	shorttitle = {Helium},
	language = {en},
	publisher = {Pergamon Press},
	author = {Ziegler, James F.},
	year = {1977},
	note = {Google-Books-ID: NCRRAAAAMAAJ},
}

@article{bragg__1905,
	title = {On the $\alpha$ particles of radium, and their loss of range in passing through various atoms and molecules},
	volume = {10},
	issn = {1941-5982, 1941-5990},
	url = {https://www.tandfonline.com/doi/full/10.1080/14786440509463378},
	doi = {10.1080/14786440509463378},
	language = {en},
	number = {57},
	urldate = {2025-09-15},
	journal = {The London, Edinburgh, and Dublin Philosophical Magazine and Journal of Science},
	author = {Bragg, W. H. and Kleeman, R.},
	month = sep,
	year = {1905},
	pages = {318--340},
}

@article{jacobs_energy_1983,
	title = {Energy spectra of neutrons produced by $\alpha$-particles in thick targets of light elements},
	volume = {10},
	issn = {0306-4549},
	url = {https://www.sciencedirect.com/science/article/pii/0306454983900038},
	doi = {10.1016/0306-4549(83)90003-8},
	number = {10},
	urldate = {2025-09-16},
	journal = {Annals of Nuclear Energy},
	author = {Jacobs, G. J. H. and Liskien, H.},
	month = jan,
	year = {1983},
	pages = {541--552},
}

@techreport{Geiger1976,
  author    = {K. W. Geiger and L. van der Zwan},
  title     = {An Evaluation of the Be($\alpha$,n) Cross Section},
  institution = {National Research Council Canada},
  number    = {NRCC-15303},
  pages     = {12},
  year      = {1976}
}

@article{Bair1979,
  author    = {J. K. Bair and J. {Gomez del Campo}},
  title     = {Neutron Yields from Alpha Particle Bombardment},
  journal   = {Nuclear Science and Engineering},
  volume    = {71},
  pages     = {18},
  year      = {1979}
}

@article{Norman1984,
  author    = {E. B. Norman and T. E. Chupp and K. T. Lesko and P. J. Grant and G. L. Woodruff},
  title     = {{22Na Production Cross Section from the 19F($\alpha$,n) Reaction}},
  journal   = {Physical Review C},
  volume    = {30},
  pages     = {1339},
  year      = {1984}
}

@techreport{Perry1981,
  author    = {R. T. Perry and W. B. Wilson},
  title     = {{Neutron Production from ($\alpha$,n) Reactions and Spontaneous Fission in ThO$_2$, UO$_2$ and (U,Pu)O$_x$ Fuels}},
  institution = {Los Alamos National Laboratory},
  number    = {LA-8869-MS},
  year      = {1981}
}

@techreport{Woosley1975,
  author    = {S. E. Woosley and W. A. Fowler and J. A. Holmes and B. A. Zimmerman},
  title     = {OAP-422},
  year      = {1975}
}
\end{document}